\documentclass[twocolumn]{aastex631}
\newcommand{\Ha}{\hbox{{\rm H}\kern 0.1em$\alpha$}}
\newcommand{\Hb}{\hbox{{\rm H}\kern 0.1em$\beta$}}
\newcommand{\MgII}{\hbox{{\rm Mg}\kern 0.1em{\sc ii}}}
\newcommand{\CIV}{\hbox{{\rm C}\kern 0.1em{\sc iv}}}
\newcommand{\NeV}{\hbox{[{\rm Ne}\kern 0.1em{\sc v}]}}
\newcommand{\OII}{\hbox{[{\rm O}\kern 0.1em{\sc ii}]}}
\newcommand{\NeIII}{\hbox{[{\rm Ne}\kern 0.1em{\sc iii}]}}
\newcommand{\OIII}{\hbox{[{\rm O}\kern 0.1em{\sc iii}]}}
\newcommand{\NII}{\hbox{[{\rm N}\kern 0.1em{\sc ii}]}}
\newcommand{\SII}{\hbox{[{\rm S}\kern 0.1em{\sc ii}]}}

\newcommand{\lmass}{$\log\mathrm{M\!_\star/M}_{\odot}$}

\newcommand{\uvnir}{[0.25$-$1 $\mu$m]~}
\newcommand{\nearmidir}{[1$-$3 $\mu$m]~}

\usepackage{amsmath,amstext,natbib,longtable,verbatim,float,graphicx,color, longtable, soul, enumitem} % ,appendix}   

\definecolor{citeRGB}{rgb}{0,0.1,0.7}
\begin{document}

\title{A Comprehensive Photometric Selection of `Little Red Dots' in MIRI Fields: An IR-Bright LRD at $z=3.1386$ with Warm Dust Emission}

\author[0000-0001-6813-875X]{Guillermo Barro}
\affiliation{University of the Pacific, Stockton, CA 90340 USA}

\author[0000-0003-4528-5639]{Pablo G. P\'erez-Gonz\'alez}
\affiliation{Centro de Astrobiolog\'{\i}a (CAB), CSIC-INTA, Ctra. de Ajalvir km 4, Torrej\'on de Ardoz, E-28850, Madrid, Spain}

\author[0000-0002-8360-3880]{Dale D. Kocevski}
\affiliation{Department of Physics and Astronomy, Colby College, Waterville, ME 04901, USA}

\author[0000-0001-8688-2443]{Elizabeth J.\ McGrath}
\affiliation{Department of Physics and Astronomy, Colby College, Waterville, ME 04901, USA}

\author[0000-0002-9393-6507]{Gene C. K. Leung}
\affiliation{Department of Astronomy, The University of Texas at Austin, Austin, TX 78712, USA}
\affiliation{MIT Kavli Institute for Astrophysics and Space Research, 77 Massachusetts Ave., Cambridge, MA 02139, USA}

\author[0000-0002-3736-476X]{Fergus Cullen}
\affiliation{Institute for Astronomy, University of Edinburgh, Royal Observatory, Edinburgh, EH9 3HJ, UK}

\author[0000-0002-1404-5950]{James S. Dunlop}
\affiliation{Institute for Astronomy, University of Edinburgh, Royal Observatory, Edinburgh, EH9 3HJ, UK}

\author[0000-0001-7782-7071]{Richard S. Ellis}
\affiliation{Department of Physics \& Astronomy, University College London, London, WC1E 6BT, UK}

\author[0000-0001-8519-1130]{Steven L. Finkelstein}
\affiliation{Department of Astronomy, The University of Texas at Austin, Austin, TX 78712, USA}

\author[0000-0001-9440-8872]{Norman A. Grogin}
\affiliation{Space Telescope Science Institute, 3700 San Martin Drive, Baltimore, MD 21218, USA}

\author[0000-0002-8096-2837]{Garth Illingworth}
\affiliation{Department of Astronomy and Astrophysics, UCO/Lick Observatory, University of California, Santa Cruz, CA 95064, USA}

\author[0000-0001-9187-3605]{Jeyhan S. Kartaltepe}
\affiliation{Laboratory for Multiwavelength Astrophysics, School of Physics and Astronomy, Rochester Institute of Technology, 84 Lomb Memorial Drive, Rochester, NY 14623, USA}

\author[0000-0002-6610-2048]{Anton M. Koekemoer}
\affiliation{Space Telescope Science Institute, 3700 San Martin Drive, Baltimore, MD 21218, USA}

\author[0000-0003-1581-7825]{Ray A. Lucas}
\affiliation{Space Telescope Science Institute, 3700 San Martin Drive, Baltimore, MD 21218, USA}

\author{Ross J. McLure}
\affiliation{Institute for Astronomy, University of Edinburgh, Royal Observatory, Edinburgh, EH9 3HJ, UK}

\author[0000-0001-8835-7722]{Guang Yang}
\affiliation{Nanjing Institute of Astronomical Optics \& Technology, Chinese Academy of Sciences, Nanjing 210042, China}

\shorttitle{Gotta Catch 'Em All}

\begin{abstract} 

JWST has revealed a population of compact ``Little Red Dots" (LRDs) at $z\gtrsim4$, with red rest-frame optical and blue UV colors. These objects are likely compact dusty starbursts or heavily reddened AGNs, playing a pivotal role in early black hole growth, dust production, and stellar assembly. We introduce a new photometric selection to identify LRDs over a broad range in redshifts and rest-frame UV-to-NIR colors enabling a more complete census of the population. This method identifies 248 LRDs with F444W\,$<27$~mag over 263 arcmin$^2$ in the JADES, PRIMER-COSMOS, and UDS fields with MIRI coverage, increasing the number density by $\times$1.7 compared to previous samples, suggesting that previous census were underestimated. Most LRDs are detected in MIRI/F770W but only 7\% (17) are detected in F1800W.  We use MIRI-based rest-frame [1$-$3 $\mu$m] colors to trace dust emission. F1800W-detected LRDs have a median [1$-$3 $\mu$m]\,$=1.5$~mag, with a broad scatter indicative of diverse dust emission properties. About 20\% exhibit colors [1$-$3 $\mu$m]\,$<1$~mag consistent with negligible dust emission, but the majority show significant dust emission at 3~$\mu$m (f\,$^{\rm dust}_{3\mu m}\lesssim0.8$) from the galaxy ISM or a hot-dust-deficient AGN torus.  A correlation between bluer UV-to-NIR colors and stronger IR emission suggests that the bluest LRDs may resemble unobscured QSOs. We report a LRD at $z_{\rm spec}=3.1386$, detected in MIRI, Spitzer/MIPS, and Herschel/PACS. Its IR SED rises steeply at $\lambda_{\rm rest}>6\,\mu$m and peaks near $\sim40\,\mu$m, providing the first direct evidence of warm dust emission (T$=50–100$~K) in a LRD.

\end{abstract}

\keywords{galaxies: spectroscopy --- galaxies:  high-redshift}

\section{Introduction}
\label{s:intro}

One of the most surprising results of the first few years of the James Webb Space Telescope (JWST) has been the discovery of a nearly ubiquitous population of compact, red sources at high-redshift that have come to be known as ``Little Red Dots" (LRDs; \citealt{matthee24}). These intriguing objects are characterized by their point-like morphology, red rest-frame optical to near-infrared colors 0.4$-$1~$\mu$m, blue rest-frame UV colors 0.15$-$0.4~$\mu$m \citep{labbe23,barro24,greene23,akins24} and a broad redshift distribution from $z\gtrsim3-9$ \citep{kocevski24,kokorev24,pg24a}. LRDs are remarkably common, with volume densities $\sim$10$^{-5}$~Mpc$^{-3}$ at $M_{\rm 0.25~\mu m}=-20$ to -17~mag, comprising a few percent of the galaxy population at these epochs \citep{barro24,labbe24,greene23}. However, despite the considerable attention and follow-up observations, the nature of their exotic properties is still heavily debated.

NIRSpec spectroscopic follow-up of LRDs selected from NIRCam colors has revealed ubiquitous emission lines, confirming their high redshift nature \citep{kocevski24,furtak23,killi23,matthee24}, and finding a surprisingly large fraction of Balmer lines with narrow plus broad components, showing widths between FWHM$=$1200$-$4000~km~s$^{-1}$ \citep{greene23,taylor24,wang24_lrd}. Such broad lines are typically indicative of direct line-of-sight into the broad line region of a low obscuration QSO, and therefore suggest the presence of Active Galactic Nuclei (AGN) in their cores. However, this interpretation appears at odds with their red optical spectral energy distribution (SEDs) which suggests a large dust attenuation. Furthermore, their peculiar blue+red colors are different from typical AGN SEDs. At best, they resemble those of heavily obscured AGNs with a small percentage of UV light scattered by the torus making up the blue color \citep{kocevski23,labbe24}. The presence of a hot dust obscuring torus with typical T$\sim$1000~K would imply a significant amount of re-emission in the near-to-far infrared (IR) wavelengths that should be easily detectable. However, in another surprising twist, early MIRI observations at wavelengths 5$-$25~$\mu$m have found very small detection fractions and relatively blue mid-IR colors inconsistent with large amounts of hot dust emission \citep{akins24,leung24,pg24a,williams24}. In light of those results, alternative AGN-dominated scenarios have been proposed where the UV excess and the lack of IR emission could be explained by warmer torus temperatures (hot-dust deficient AGN; e.g.; \citealt{lyu24}) in a lower-luminosity AGN hosted by a blue, low-mass galaxy host \citep{pg24a,wang24_lrd,wang24b,akins24,leung24}.

An alternative scenario to explain the modest mid-IR colors is that the continuum in that spectral range is stellar dominated, peaking around the 1.6~$\mu$m stellar bump and showing a relatively flat SED into the mid-IR (e.g., \citealt{pg24a}; \citealt{williams24}). While the broad emission lines could still be associated with an AGN, the stellar emission would dominate the continuum and the dust attenuation would originate in the galaxy-wide interstellar medium (ISM). This scenario, however, also has some complications. First, the bimodal SEDs are difficult to reconcile with typical stellar population models requiring either an extremely grey attenuation law or two distinct attenations, low and high, to fit the blue+red SED (e.g., \citealt{barro24}; \citealt{pg24a}; \citealt{kocevski24}). This configuration could describe a galaxy with a patchy dust distribution where some UV emission escapes unobscured. However, additional constraints from the lack of sub-mm detections in stacks of LRDs (\citealt{labbe24}; \citealt{akins24}) imply that the dust must be very concentrated and have a high effective temperature to shift the peak of the IR re-emission to shorter wavelengths (\citealt{pg24a}; \citealt{akins24}; \citealt{casey24}). The ultra-compact sizes of LRDs also place restrictive upper limits to the galaxy radius r$\lesssim$100~pc (\citealt{baggen23}; \citealt{akins24}), which combined with the relatively large stellar masses inferred from the SED-fitting lead to large stellar densities close to the expectations for halo baryon abundances in $\Lambda$CDM (\citealt{boylan23}; \citealt{dekel23}; \citealt{leung24}).

The unorthodox nature and seemingly identical SED fits obtained with the different SED models make it difficult to distinguish between galaxy and AGN-dominated scenarios or some combination of both. Recently, \citet{wang24b} found both stellar and AGN signatures in the form of Balmer breaks and broad Balmer emission lines pointing toward a complex co-existence between galaxy host and AGN. However, the continuum features could also be associated with large fractions of dense gas around the AGN \citep{inayoshi24}.  

Beyond SED degeneracies, another major challenge in characterizing this complex population is defining what exactly constitutes an LRD. Photometric and spectroscopic selections highlight different features (red colors vs. broad emission lines), which may not be present in all sources. The potential that LRDs represent a rare type of AGN further complicates the issue, as JWST spectroscopy is revealing an unexpectedly large population of broad-line AGNs at high redshift (\citealt{kocevski23}; \citealt{harikane23a}; \citealt{maiolino23}; \citealt{larson23}; \citealt{taylor24}), though not all of them are LRDs (e.g.,\citealt{taylor24}; \citealt{hainline24}). Additionally, large spectroscopic follow-ups of photometric samples often focus on a small subset of LRDs, typically the reddest sources (e.g., \citealt{degraaff24}), while photometric selections sometimes use different color and morphological thresholds. This can lead to biased or incomplete samples that fail to capture the full diversity of LRDs. For example, \citet{pg24a} recently showed that LRDs span a relatively broad range of rest-frame UV-to-NIR colors, which may correlate with their mid-IR properties. Likewise, \citet{kocevski24} demonstrated that using different bands to probe the rest-frame UV and optical can extend the redshift range to lower and higher redshifts ($3<z<9$), as opposed to the typical $z\gtrsim5$ seen in earlier selections.

In this paper, we present a new photometric selection designed to capture the broad range in redshift and rest-frame UV-to-NIR colors of the LRDs, and provide a more complete picture of their properties. We identify a sample of 248 LRDs over the 263 arcmin$^{2}$ of the JADES, PRIMER-COSMOS, and UDS fields with MIRI coverage. The latter allows us to extend the UV-to-optical SED coverage from NIRCam to the near and mid-IR. These regions are key to identifying bluer LRDs without emission line contamination and probing the amount of dust IR emission. The paper is organized as follows. \S~\ref{s:data} describes the photometric datasets used. \S~\ref{s:sample} describes the new photometric selection compared to previous samples, discusses the MIRI photometric properties of the sample, and overall number density. \S~\ref{s:properties} analyzes the redshifts, rest-frame UV-to-NIR colors, luminosities, and IR emission of the sample. \S~\ref{s:lowz_LRDs} presents a detailed analysis of two LRDs at the lowest redshift of the sample using constraints from MIRI to probe deeper into the rest-frame mid-IR ($\lambda\sim5\mu$m) and discuss galaxy vs. AGN SED modeling scenarios. 

Throughout the paper, we assume a flat cosmology with $\mathrm{\Omega_M\, =\, 0.3,\, \Omega_{\Lambda}\, =\, 0.7}$, and a Hubble constant $\mathrm{H_0\, =\, 70\, km\,s^{-1} Mpc^{-1}}$. We use AB magnitudes \citep{oke83}.

\section{Data}
\label{s:data}
The primary goal of this paper is to review the photometric selection of LRDs aiming to obtain more complete samples representative of their variety in rest-frame UV-to-NIR colors.  Probing the rest-frame NIR at $\lambda\sim1\mu$m with JWST requires MIRI data. Therefore we focus our analysis in JWST deep fields with a relatively large coverage of public MIRI imaging. Specifically, we use the two fields of the Public Release IMaging for Extragalactic Research (PRIMER) survey \citep{dunlop21} and the GOODS-S field of the JWST Advanced Deep Extragalactic Survey (JADES; \citealt{eisenstein23}), focusing on the regions with MIRI coverage.

\subsection{Optical and NIR photometry}
Both PRIMER fields have NIRCam photometry in 8 bands (F090W, F115W, F150W, F200W, F277W, F356W, F444W, and F410M). JADES has the same NIRCam bands but, in addition, it also overlaps partially with the FRESCO \citep{oesch23} and JEMS \citep{williams23} surveys which provide NIRCam medium band photometry in up to 6 additional filters (F182M, F210M, F335M, F430M, F460M, F480M). The JADES has ancillary HST/ACS and WFC3 photometry in 7 bands (F435W, F606W, F814W, F105W, F125W, F140W, and F160W) while PRIMER has the same WCF3 bands but only 2 ACS bands, F606W and F814W, where the HST data in these fields come from several programs, among the largest of which are GOODS \citep{giavalisco04} and CANDELS \citep{candelsgro,candelskoe}.

For source extraction and photometry in the PRIMER fields, we use the latest version (v7.2) of the publicly available catalogs from the DAWN JWST Archive (DJA). The DJA photometry retrieves the original JWST/NIRCam level-2 products and reduces them with grizli \citep{brammer_grizli}. The source detection is performed with SEP \citep{SEP}, a Python version of SExtractor (\citealt{sex}). Photometry is measured in 0.5'' circular apertures and corrected to total \citet{kron80} using aperture corrections (see \citealt{valentino23} and Brammer in prep for more details). For the JADES field, we use the public, multi-band photometric catalogs from DR2 \citep{eisenstein24}.

\subsection{Mid-IR photometry}
The PRIMER survey has MIRI coverage in two bands, F770W and F1800W covering a fraction of the NIRCam mosaic with total areas of 105 and 123 arcmin$^{2}$ in UDS and COSMOS, respectively. In GOODS-S, the Systematic Mid-infrared Instrument Legacy Extragalactic Survey (SMILES; \citealt{rieke24}, \citealt{albers24}) covered the central 36 arcmin$^{2}$ overlapping with the UDF in 8 MIRI bands covering from 5 to 26 microns. We use the MIRI observations from PRIMER and SMILES reduced and processed following the procedures described in detail in \citet{pg24a} and \citet{leung24}. The photometry for the LRDs was computed using circular apertures following the same procedures as in the aforementioned works. Briefly, we measured fluxes and uncertainties in several circular apertures ranging in radii between 0.2'' and 0.8'' and applied aperture corrections (see Appendix A in \citealt{pg24a}) for point-like sources to estimate the total flux. Then, for each source, we use the total magnitude from the aperture with the highest signal-to-noise that is consistent with the smaller apertures at a 1$\sigma$ level.

After the photometric selection of LRDs described in the following sections, we used the combined NIRCam+MIRI SEDs to recompute the initial redshift estimates from the JDA and JADES catalogs using \texttt{EAZYpy} \citep{eazy} with the addition of the LRD-like template {\it j0647agn+torus} \citep{killi23} template. We also derive the rest-frame colors used throughout the paper by fitting the SED at the photometric redshift with the Prospector-AGN code for LRDs described in \S~\ref{s:fittingcodes}.

\section{Photometric selection of LRDs}
\label{s:sample}

\subsection{Selection criteria}
\label{s:criteria}
\begin{figure*}%[htp!]%[ht!]
\centering
\includegraphics[width=18.5cm,angle=0]{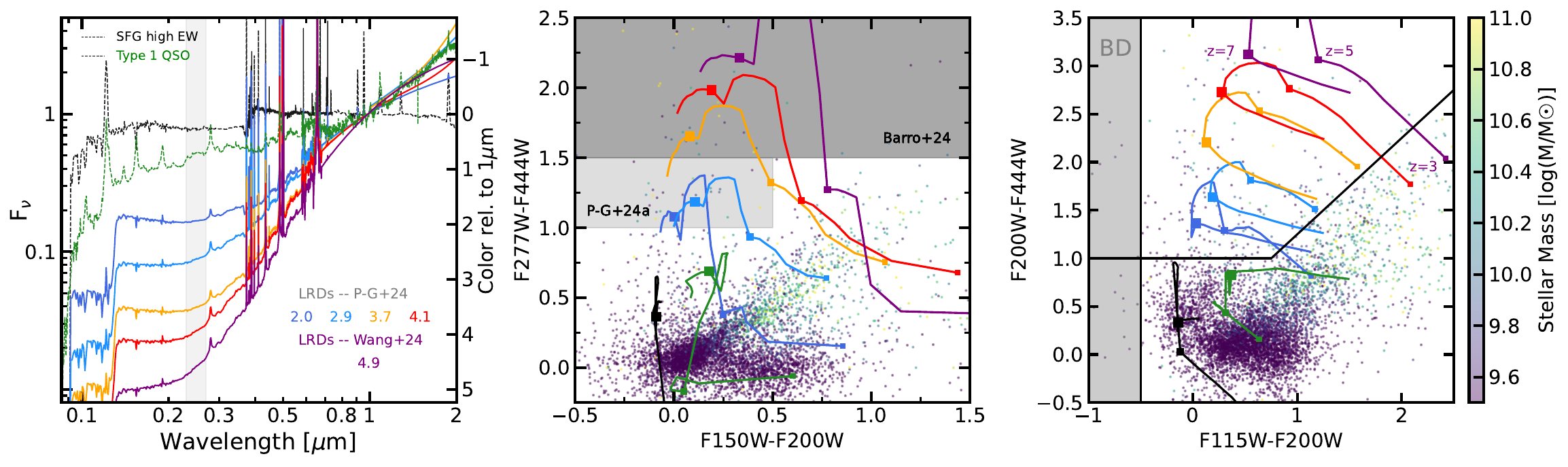}
\caption{NIRCam color-color diagrams illustrating photometric selection methods for LRDs and their typical SEDs. {\it Left:} Best-fit SED models for 5 LRDs (solid lines) drawn from \citet{pg24a} and \citet{wang24_lrd}, covering a representative range of UV-to-NIR colors indicated in the text. The black and green dashed lines show typical SEDs for a young, low-extinction star-forming galaxy with extreme emission lines (EW$\sim$1000\AA) and a type I QSO, respectively. The next panels display the color-redshift tracks of these models ranging from $z=3$ to $z=9$, as indicated by squares of increasing size. {\it Center:} Color-color selection diagram. The dark and light grey regions indicate the selection thresholds for LRDs used in \citet{barro24} and \citet{pg24a}. The density plot shows the bulk of the galaxy sample color-coded by stellar mass. The color-redshift tracks suggest that this method tends to miss LRDs at low redshift ($z\lesssim5$) and those with bluer UV-to-NIR colors (\uvnir$<3$~mag). {\it Right:} New color-color selection for LRDs. This method provides a more complete sample by selecting candidates within the region that encloses the color tracks of the representative LRD models from $z=3$ to $z=9$. The grey shaded area indicates the threshold used to reject brown dwarf contaminants.}
\label{fig:selection_criteria}
\end{figure*}

Photometric selections of LRDs rely on observed or rest-frame colors to identify their distinctive bimodal SEDs, defined by red rest-frame optical and blue UV colors. However, LRDs display significant color diversity beyond these key features. For example, while their UV slopes in f$_{\nu}$ are generally flat, their optical slopes exhibit
more diversity. Recently, \citet{pg24a} demonstrated that the UV-to-NIR colors of LRDs span nearly 2~mag, ranging from \uvnir$=1.5$ to 3.5~mag. The UV-to-NIR color, which requires MIRI to probe the rest-frame 1~$\mu$m range, provides a reliable metric for capturing the diversity of LRD types. Using a long wavelength baseline reduces
sensitivity to strong optical emission lines, which can be prominent in LRDs and artificially enhance NIRCam-based colors.

The left panel of Figure~\ref{fig:selection_criteria} illustrates the range in UV-to-NIR colors of the LRDs, showing the rest-frame SED models of four LRDs drawn from \citet{pg24a}, spanning a color range of [0.25–1~$\mu$m] from $\sim$2 to 3.5~mag. We also include a fifth model based on the LRD from \citet{wang24_lrd}, one of the reddest reported LRDs, with a color of \uvnir$\sim5$~mag. Additionally, two non-LRD models with bluer colors are shown: a young ($<100$~Myr), low-extinction star-forming galaxy (SFG) with prominent emission lines (black) and a type I QSO template (green) from \citet{vandenberk01}. The SFG is typically blue across all wavelengths, but emission lines can boost the F444W flux at certain redshifts, resulting in red NIRCam colors similar to those of LRDs. However, these emission lines do not affect the UV-to-NIR color, which remains close to \uvnir$\sim0$, providing a clear method for distinguishing emission-line SFG interlopers. The type I QSO is slightly redder than the young SFG but significantly bluer than typical LRD templates, with \uvnir$<1$mag. The color-redshift tracks of the SFG and type I QSO are shown in subsequent panels of Figure\ref{fig:selection_criteria} to highlight regions where these objects could contaminate the LRD sample.

The central panel of Figure~\ref{fig:selection_criteria} illustrates a typical color-color diagram used to identify LRDs (e.g., \citealt{barro24}; \citealt{pg24a}; \citealt{akins24}). In this diagram, the F150W$-$F200W and F277W$-$F444W colors probe the flat UV and steep optical slopes, respectively. The light-grey shaded region represents the color thresholds from \citet{pg24a}, requiring blue UV colors (F150W$-$F200W$<0.5$~mag) and red optical colors (F277W$-$F444W$>1$~mag). The darker grey area shows a stricter criterion from \citet{barro24} (also \citealt{akins24}), targeting the reddest LRDs with F277W-F444W$>1.5$~mag. This more restrictive selection reduces contamination from SFGs with extreme emission lines. Notably, this selection does not include a secondary threshold in F150W$-$F200W because other potential red galaxy contaminants predominantly exhibit bluer colors (F277W$-$F444W$<1.5$~mag), as shown by the general trend for massive, dusty, or evolved galaxies in the scatter plot, color-coded by stellar mass.

The colored lines indicate the color-redshift tracks for the LRD models shown in the left panel. These tracks demonstrate that while the color thresholds efficiently recover the reddest LRDs at $z\gtrsim5$, they systematically miss LRDs at lower redshifts or with bluer colors. Low-redshift LRDs are missed because they overlap with the locus of red, dusty, or evolved galaxies on the right side of the diagram. This occurs because F150W-F200W probes the red rest-optical rather than the blue rest-UV at $z\lesssim5$. Similarly, some LRDs with bluer UV-to-NIR colors can be missed by the F277W-F444W threshold if they lack prominent emission lines to boost their NIRCam colors. This effect is more noticeable between $z=5$ and 7 (squares in the color tracks), where the \OIII\ line shifts out of F277W and H$\alpha$ shifts into F444W. This results in a strong color spike that is visible in all tracks but most prominent ($\gtrsim1$~mag) in the track of the bluest LRD model, which corresponds to JADES-204851. This object, which exhibits broad, high-EW emission lines detected in the NIRCam grism (\citealt{matthee24}), shows photometric excesses in NIRCam, as noted in \citet{pg24a}.

A way forward to address some of the issues in the photometric selection of LRDs is to use multiple observed colors to trace the rest-frame UV and optical across redshift, as demonstrated in \citet{greene23} and \citet{kokorev24} (e.g., using the {\it red1} and {\it red2} thresholds). Similarly, \citet{kocevski24} employed a selection based on the rest-frame UV and optical slopes computed from different bandpasses in four redshift bins. Here, we propose a complementary approach based on a single color-color diagram that uses slightly different bands and thresholds, informed by the color-redshift tracks of the LRD models. The right panel of Figure~\ref{fig:selection_criteria} illustrates this selection diagram, with the thresholds indicated by black lines at:

\begin{enumerate}[label=(\roman*),leftmargin=0.5cm]
    \item $ F200W-F444W>1 \, $
    \item $ (F200W-F444W) > (F115W-F200W) + 0.25 \,$
    \item $ F115W-F200W > - 0.5 \,$
    \item $ F444W< 27 \, $
\end{enumerate}

Relative to the previous criterion, the new selection shifts to the closest shorter wavelength band in the first filter of both color pairs. This provides a longer wavelength baseline, allowing for a better trace of the SED shape. Using F200W as the pivot point improves the sampling of the rest-frame UV, as this filter is blueward of 4000~\AA\ at $z>4.5$, while F277W probes the rest-frame optical over most of the redshift range up to $z\sim7$. F200W is also less affected by emission lines (with \OIII\ entering the filter at $z=3$) compared to F277W, which probes the \OIII\ or H$\alpha$ lines between $3<z\lesssim5.5$. As a result, the new color combination mitigates the strength of the emission line-driven spikes in the color-redshift tracks at $z>5$ and provides a smoother evolution that runs nearly parallel to the x-axis, regardless of strong emission lines—although a small spike remains due to H$\alpha$ entering F444W.

Another key difference is that the new selection uses a diagonal threshold with a unity slope and a small offset (F200W$-$F444W $>$ F115W$-$F200W + 0.25), rather than just a limit on the short-wavelength color (F115W$-$F200W$>1$~mag). As shown in Figure~\ref{fig:selection_criteria}, this threshold runs nearly parallel to the locus of massive, dusty/evolved galaxies, which are uniformly red across their entire SED. The shift to the left in the color tracks at $z>3$ is driven by F115W-F200W probing the rest-frame UV, revealing the characteristic blue emission that distinguishes LRDs from traditional EROs. The latter can, in extreme cases, appear similarly red and compact in the rest-optical, but they lack the blue UV emission seen in LRDs, as shown in \citet{zavala23} and more recently in \citet{kokorev24b}. Compared to the F277W-based selection, the new method significantly reduces the overlap between the color-redshift tracks and the locus of massive red galaxies, facilitating the identification of new LRDs at lower redshift. Additionally, a compactness threshold helps minimize contamination from resolved massive red galaxies (see next section). Finally, the F115W$-$F200W color provides an effective means to eliminate brown dwarf contaminants, which typically have steeper UV slopes (grey shaded area) than LRDs (\citealt{greene23, hainline23}). However, a less desirable effect is that the F115W$-$F200W color becomes very red at $z\gtrsim8$ as F115W approaches the Lyman break, causing the color-redshift tracks to exit the selection region, thereby reducing the completeness of the selection at high redshifts.

\begin{figure}%[htp!]%[ht!]
\centering
\includegraphics[width=7.5cm,angle=0]{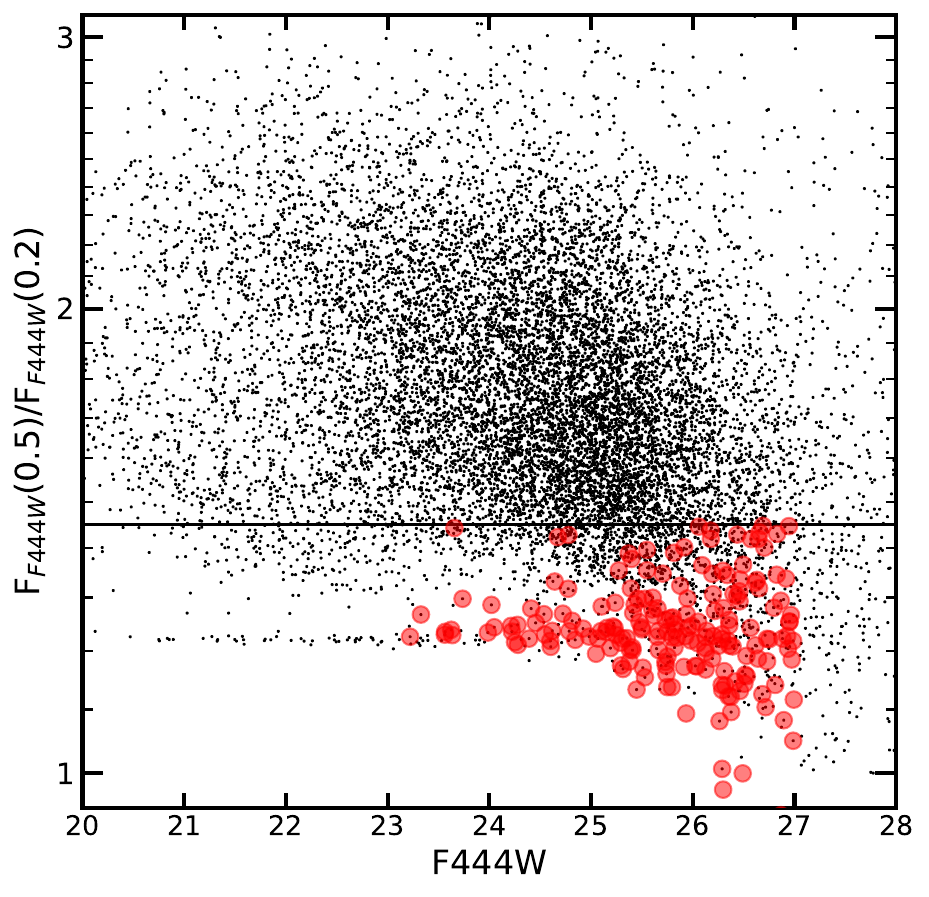}
\caption{ The flux ratio within apertures of radius r = 0.5'' and 0.2'' in F444W for the full photometric sample in PRIMER and JADES is shown by black dots. The unresolved stellar locus appears as a straight line at F$_{\rm F444W}$(0.5'')/F$_{\rm F444W}$(0.2'')~$\sim1.2$. The black line marks the compactness threshold used to identify LRDs (red circles).}
\label{fig:compactness}
\end{figure}

\begin{figure*}%[htp!]%[ht!]
\centering
\includegraphics[width=8.5cm,angle=0]{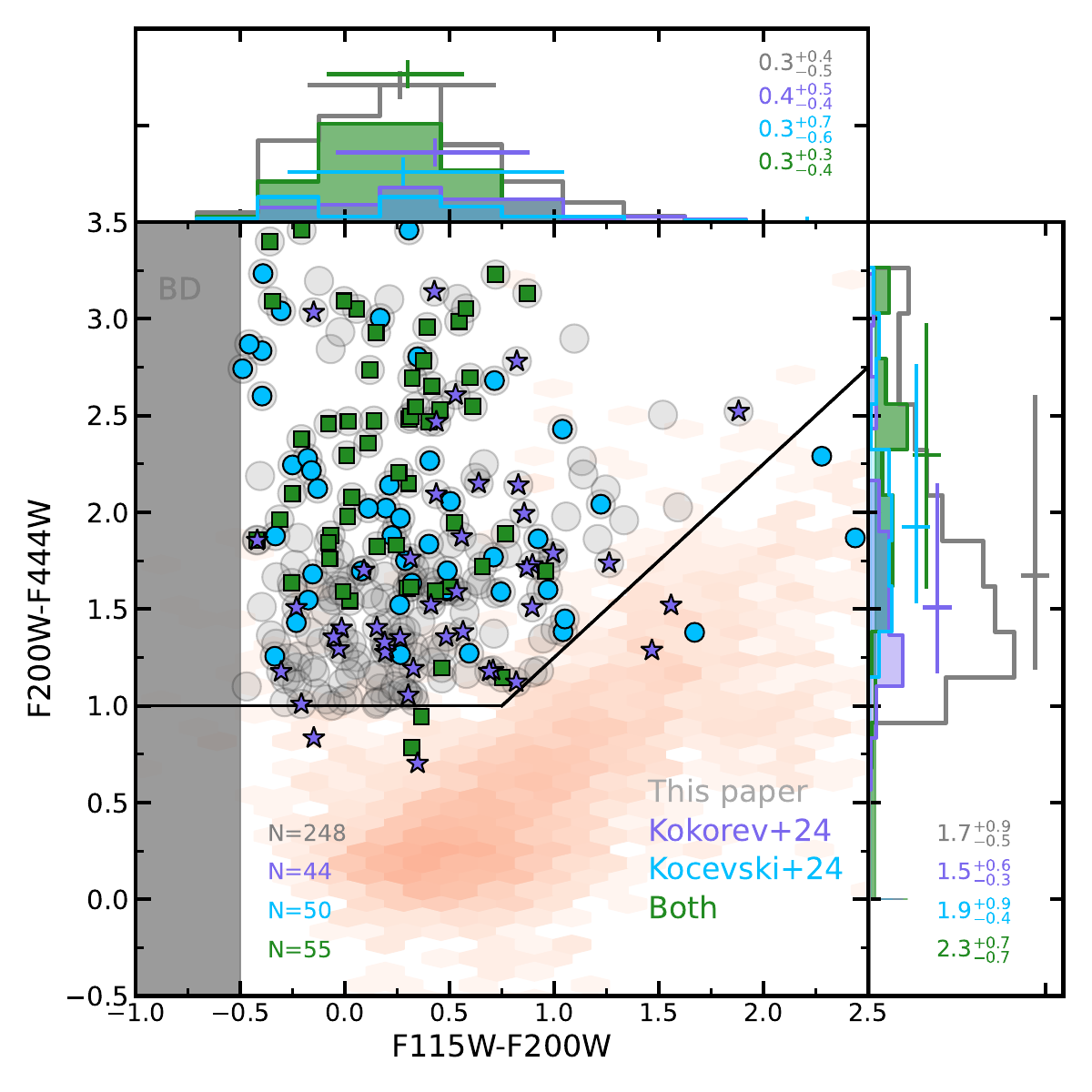}
\includegraphics[width=8.5cm,angle=0]{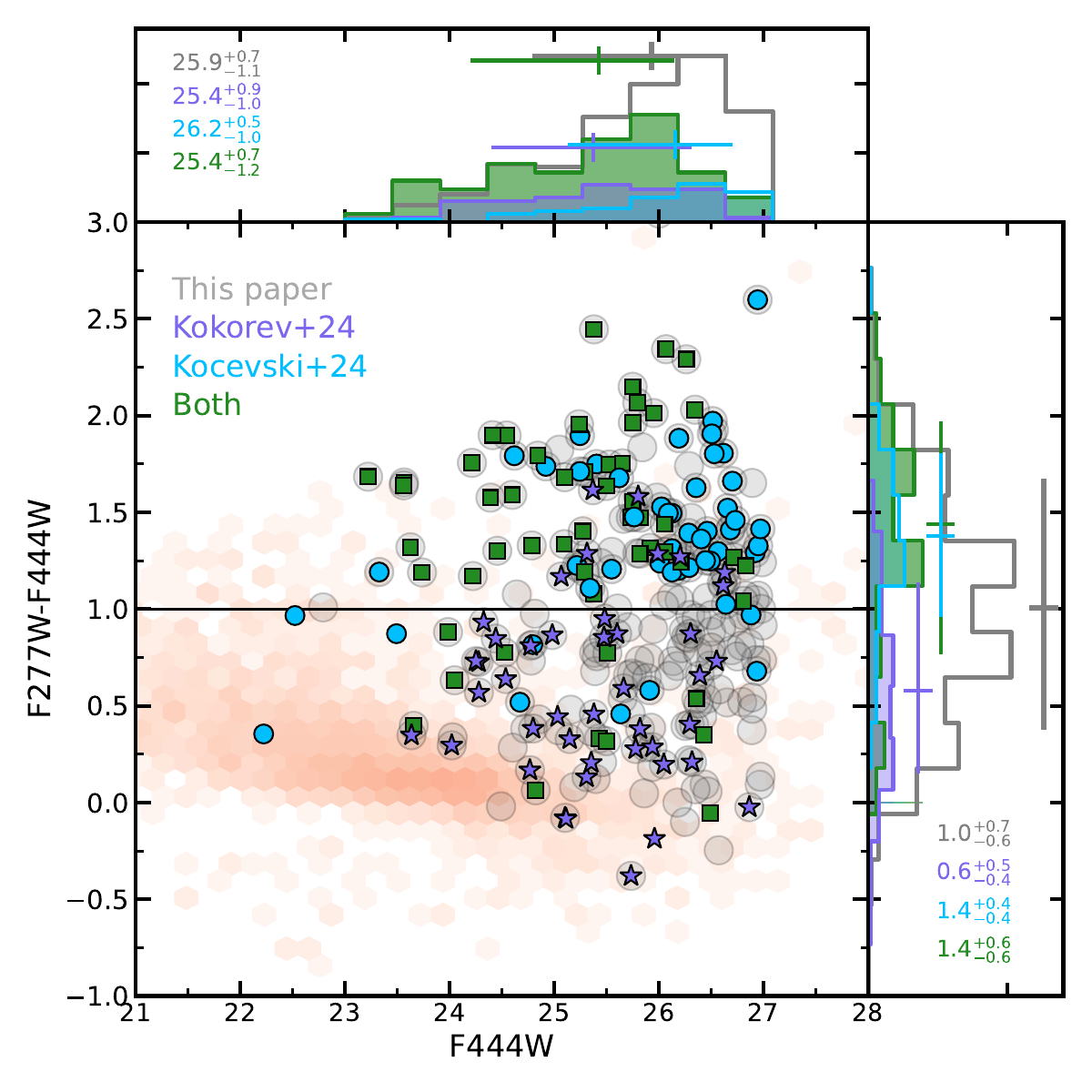}
\caption{{\it Left:} New color-color selection diagram (as in the right panel of Figure~\ref{fig:selection_criteria}) comparing to other LRD samples from the literature in the same fields. The grey circles indicate our sample. The purple stars, blue circles, and green squares show the LRDs identified only in KOV24, KOI24, and those in common between the two. The histograms on the top and side indicate the overall distribution of the different subsets, as well as their median and percentiles. {\it Right:} F277W$-$F444W color vs. F444W magnitude for the same LRDs in the left panel. The black line indicates the LRD selection threshold of \citet{pg24a}. Despite having a similar number of LRDs, the overlap between KOV24 and KOI24 is only $\sim$50\%. The agreement is better for red and bright LRDs, while KOV24 tends to miss some of the red but faint, and KOI24 misses preferentially blue LRDs with F277W$-$F444W$<1$. The new method identifies nearly all the LRDs in the previous samples and an additional 40\%, with the new LRDs being preferentially blue and faint.}
\label{fig:selection_othersamples}
\end{figure*}

The F200W$-$F444W$>1$~mag threshold was chosen to cover all LRD color-redshift tracks, including the bluest model while excluding the SFG and type I QSO tracks. This choice balances LRD coverage with avoiding denser regions of the diagram, where contamination from strong emission lines could increase. In the following sections, we show that using the rest-frame UV-to-NIR color derived from MIRI/F770W, instead of F444W, further reduces the influence of emission-line contaminants with red F200W$-$F444W$>1$~mag but blue \uvnir$\sim0$~mag colors. The steep optical SED of LRDs implies that using a color threshold of F200W$-$F444W$>1$ instead of F277W$-$F444W$>1$ can include some intrinsically bluer LRDs with SEDs between the type I QSO and the bluest LRD template, i.e., \uvnir$=1$ to 2~mag. This color space is particularly interesting for exploring potential connections between the blue, unobscured AGN population and LRDs. While the nature of LRDs is still debated, one possibility is that the blue UV light comes from scattered AGN radiation, resembling a type I QSO SED, while the red optical emission arises from an obscured accretion disk (e.g., \citealt{kocevski23}; \citealt{greene23}; \citealt{labbe24}). In this scenario, the relative luminosities of the scattered and obscured components could shift LRDs within the UV-to-NIR color range, from the reddest models toward the bluer, QSO-like template.

\subsection{Compactness criterion}

The initial LRD papers focused on the photometric identification of sources with blue+red SEDs, without any size constraints. The compactness and borderline unresolved morphology of these sources (e.g., \citealt{labbe23}; \citealt{barro24}; \citealt{pg24a}) emerged as a result of the selection criteria rather than a fundamental requirement, and were interpreted as further evidence that these photometrically selected objects constituted a distinct population. However, as photometric selections have expanded to include more complete LRD samples across redshift and intrinsic color ranges, the likelihood of contamination from extended galaxies has become more relevant (e.g., \citealt{greene23}). As a result, a compactness criterion is now widely used to exclude extended objects. Recently, \citet{kocevski24} found that by adopting a size threshold based on the SExtractor half-light radius relative to the size of stars (r$_{\rm h}<1.5~r_{\rm h, star}$), approximately 20\% of the objects in the photometric selection were extended. However, those contaminants were mostly bright galaxies at lower redshifts (F444W$<24$~mag and $z<4$), which explains why earlier works targeting LRDs at $z>5$ did not require a size threshold. In this work, we apply a similar compactness criterion as \citet{greene23} and \citet{kokorev24}, using the flux ratio measured in two different apertures (F$_{\rm F444W}$(0.5'')/F$_{\rm F444W}$(0.2'')$<1.5$), as illustrated in Figure~\ref{fig:compactness}.

\begin{figure*}%[htp!]%[ht!]
\centering
\includegraphics[width=8.5cm,angle=0]{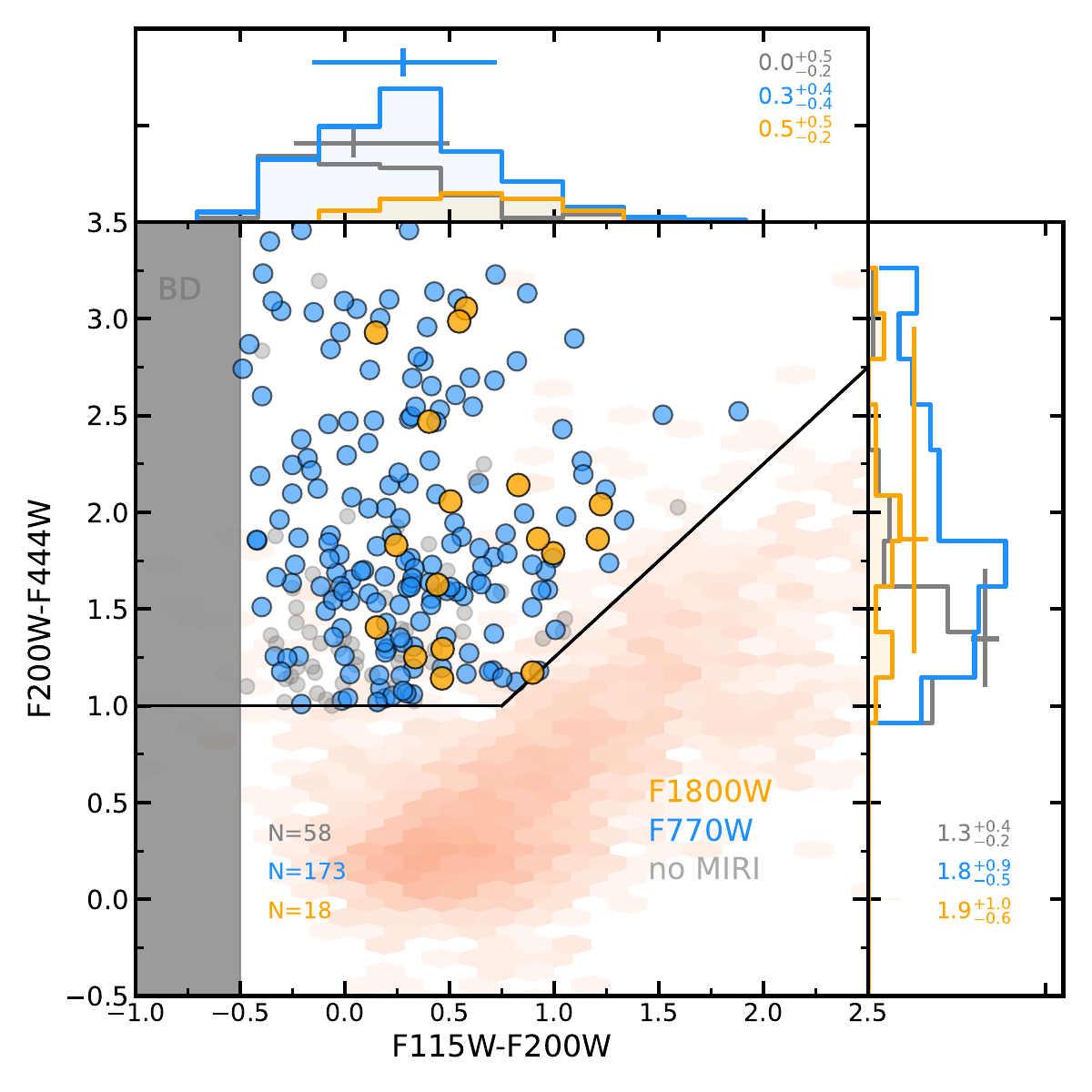}
\includegraphics[width=8.5cm,angle=0]{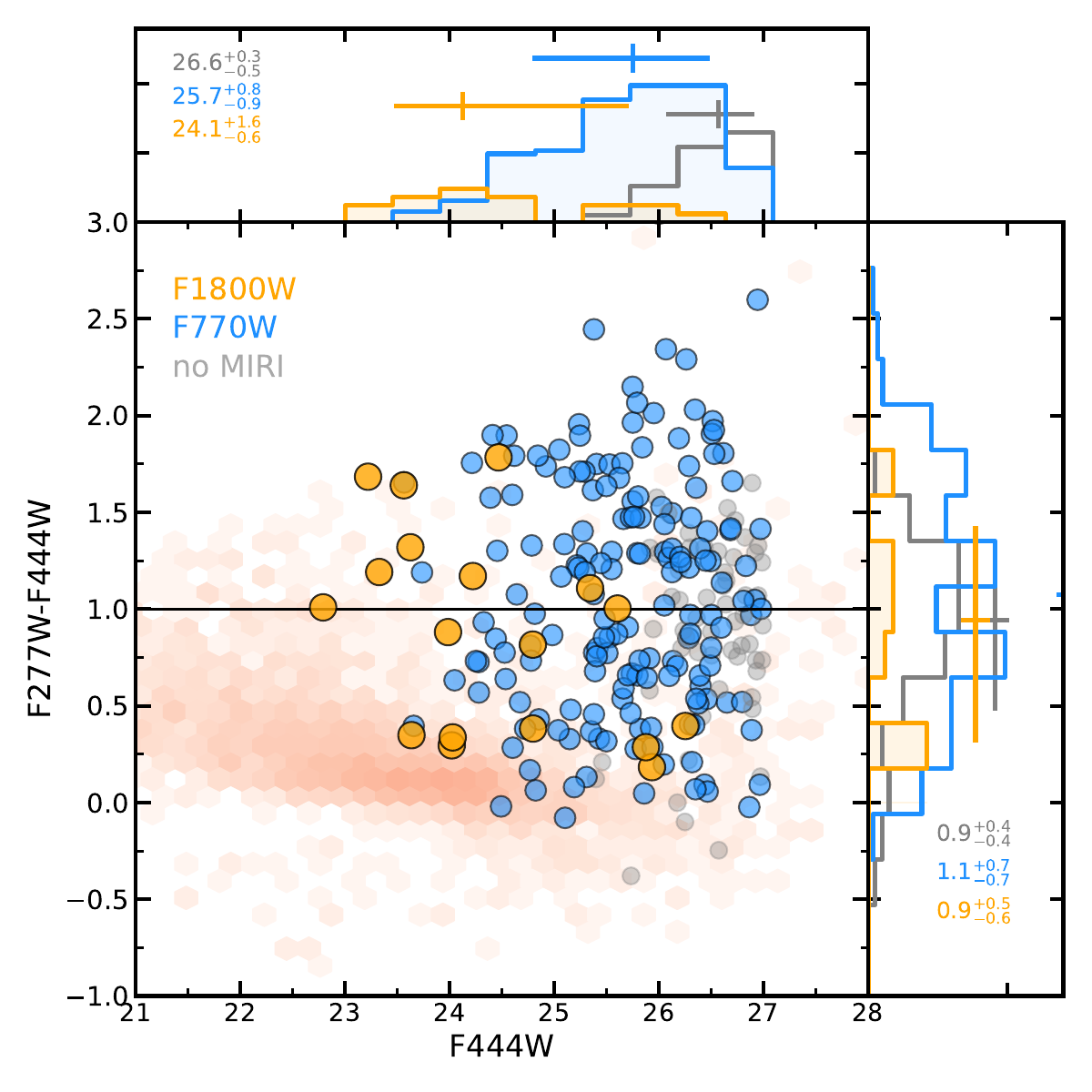}
\caption{Same color-color and color-magnitude diagrams as in Figure~\ref{fig:selection_othersamples}, but color-coded based on detection in different MIRI bands. Orange, blue, and grey circles represent LRDs detected in F1800W, F770W, and those not detected in MIRI, respectively. The majority of the LRD sample is detected in F770W, with the detection fraction primarily dependent on the F444W magnitude, ranging from 77\% to 94\% for F444W$<27$~mag and F444W$<26$~mag, respectively. In contrast, the detection fraction in F1800W is only $\sim7$\% (17), predominantly concentrated among the brightest LRDs (F444W$\lesssim24$~mag). However, there is a weak trend of F1800W detections extending to fainter LRDs (F444W$>25$~mag) with bluer F277W$-$F444W colors.}
\label{fig:selection_MIRI}
\end{figure*}

\subsection{LRD sample: Observed colors and comparison to previous samples}

We apply the photometric selection described in the previous section to identify LRDs in the regions of the PRIMER and JADES fields with MIRI coverage. Figure~\ref{fig:selection_othersamples} compares our sample to recent LRD samples from \citet{kokorev24} (hereafter KOV24) and \citet{kocevski24} (hereafter KOI24) using color-color and color-magnitude diagrams. As noted earlier, KOI24 combines different NIRCam bands as a function of redshift to identify LRDs with blue UV and red optical slopes. Their method, shown in Figure 3 of their paper, identifies the LRDs selected with a F277W$-$F444W$>1$~mag threshold (as in \citealt{pg24a} and \citealt{akins24}), but also new LRDs at lower redshifts that were missed in previous studies. Similarly, KOV24 uses the {\it red1} and {\it red2} color combinations from \citet{greene23} and \citet{labbe24}, aimed at identifying blue+red SEDs at low and high redshifts, respectively. The {\it red2} color is similar to the threshold in \citet{pg24a}, while the {\it red1} color is analogous to KOI24's choice of bands for $3<z\lesssim5$. Thus, the LRD samples from these studies should largely overlap with each other and our new selection, which uses longer baseline colors to account for band shifts with redshift.

Interestingly, Figure~\ref{fig:selection_othersamples} shows that while the KOI24 and KOV24 samples have a similar total number of sources (105 and 99, within the MIRI region and F444W$<$27 mag), there are significant differences between them. Only 55 LRDs are identified simultaneously by both methods, which means roughly 52\% and 55\% of each sample, respectively. The similarities and differences between the two are more apparent in the color-magnitude diagram on the right. The LRDs in common (green squares) tend to be preferentially red (F277W$-$F444W$>1$~mag) and bright (F444W$\lesssim26$~mag). The LRDs only in KOI24 (blue circles) are also red, but typically fainter, with F444W$\gtrsim26$~mag. In contrast, the LRDs only in KOV24 (purple stars) are preferentially bluer in F277W$-$F444W$<1$~mag, bright (F444W$\lesssim26$~mag), and at a lower redshift ($z\lesssim5$; see also \S~\ref{s:properties}). The bluer color in F277W$-$F444W is primarily due to \OIII\ contamination in F277W, which enters the filter between $z=4$ and 5.5. This excess causes the median F200W$-$F277W color to increase from 0.25~mag to 1~mag at $z\lesssim5.5$. As a result, at these redshifts, LRDs with relatively blue F200W-F444W$\sim$1 colors will have even bluer F277W-F444W colors, potentially reaching F277W$-$F444W$\sim$0~mag, as indicated by the purple stars. 

As a result of these differences, there are a total of 149 LRDs identified by either one of these works, which implies an additional 41\% and 50\% more LRDs relative to the samples in KOV24 or KOI24. It is also worth noting that despite the use of different bands as a function of redshift, 84\% of the LRDs in KOI24 satisfy the F277W$-$F444W$>1$~mag criterion from \citealt{pg24a}. The left panel of Figure~\ref{fig:selection_othersamples} shows that the bulk of the LRDs in KOV24 or KOI24 fall within the selection region of the new method. Consequently, our sample finds 93\% of the LRDs in any of the previous works and 98\% of those in common between the two. The small number of LRDs missed by our method are relatively close to the selection boundaries. Overall, the new selection finds 248 LRDs, 40\% previously unidentified, roughly a factor of $\times1.7$ more LRDs than the previous samples combined. These new LRDs tend to be fainter, with a median F444W$=26.2$~mag, and bluer with median colors F277W$-$F444W$=0.8$~mag and F200W$-$F444W$=1.4$~mag. As we will show in \S~\ref{s:restcolors}, these bluer LRDs exhibit the usual V-shaped SED of LRDs with UV-to-NIR colors larger than \uvnir$>1$~mag and therefore different from the low-mass SFGs or the Type I QSO models in the left panel of Figure~\ref{fig:selection_criteria}.

\subsection{LRD Sample: MIRI detections}

Figure~\ref{fig:selection_MIRI}  shows the same color-color and color-magnitude diagrams as Figure~\ref{fig:selection_criteria}, color-coded by MIRI detection in F770W and F1800W down to 5$\sigma$ limits of F770W$=25.7$~mag and F1800W$=23.0$~mag. Overall, 77\% of the LRD sample is detected in F770W (blue circles), with a median magnitude of F770W$=25.3$~mag. The detection fraction depends strongly on the F444W magnitude, increasing to 94\% at F444W$<26$~mag. Thus, the F770W non-detections (grey circles) are typically faint, with a median F444W$=26.6$~mag, and exhibit bluer F200W$-$F444W$=1.4$~mag and F277W$-$F444W$=0.9$~mag colors than most of the other subsets in Figure~\ref{fig:selection_othersamples}. On the other hand, the detection fraction in F1800W is much lower, at only 7\%. Most of these detections (orange circles) are concentrated on the brightest LRDs, with a median magnitude F444W$=24.1$~mag, increasing the detection fraction to nearly 70\% at F444W$<24$~mag. Nevertheless, a handful of F1800W detections have fainter magnitudes suggesting a weak negative correlation toward bluer colors. The F1800W detections at F444W$>24.5$~mag exhibit a bluer median color F277W$-$F444W$=0.4$~mag than those at F444W$<24.5$~mag, which have a redder median color of F277W$-$F444W$=1$~mag.

\begin{figure}%[htp!]%[ht!]
\centering
\includegraphics[width=8cm,angle=0]{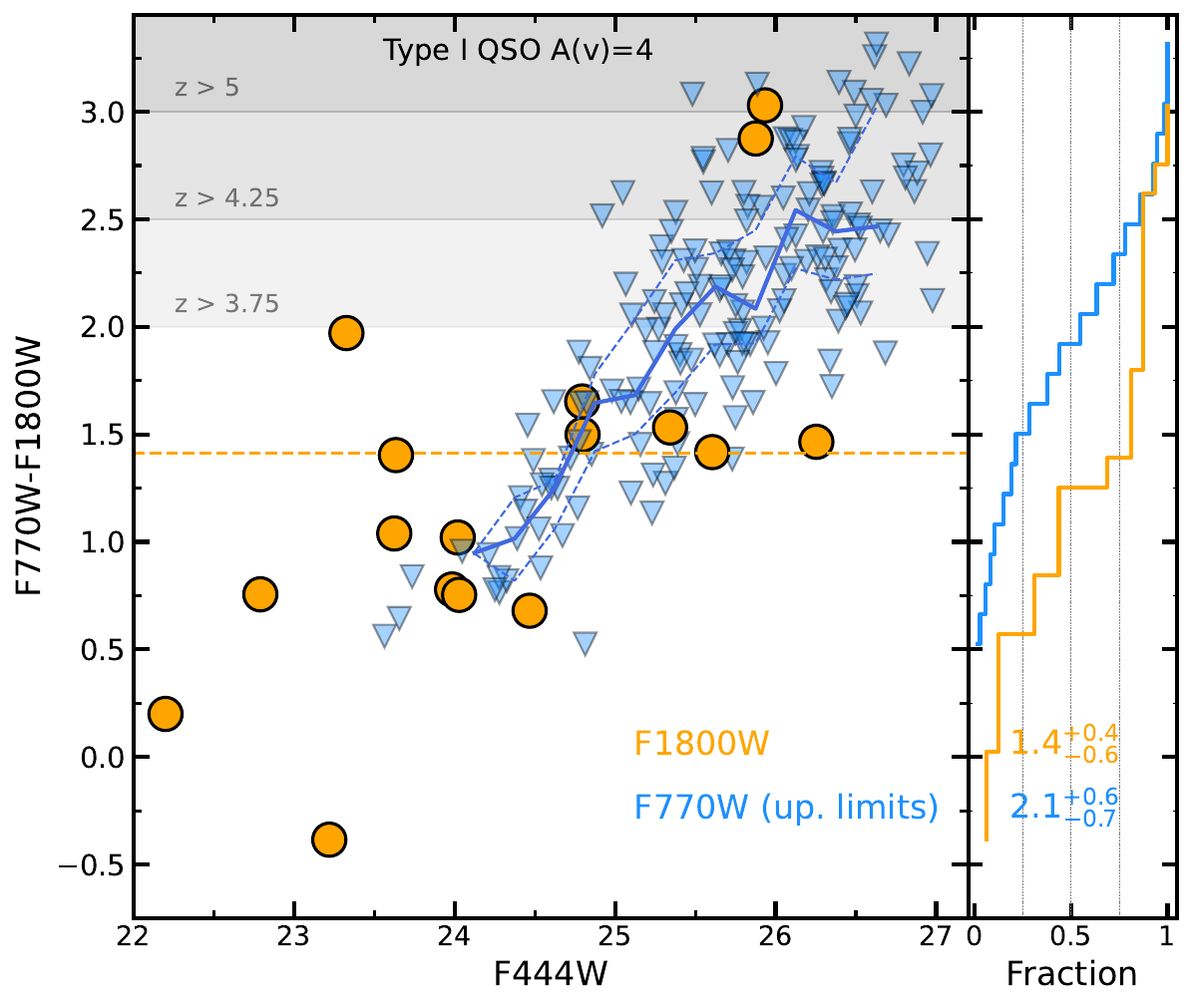}
\caption{F1800W$-$F770W color vs. F444W magnitude for MIRI-detected LRDs. Orange circles represent F1800W detections, and blue triangles indicate upper limits for LRDs detected only in F770W. The blue lines show the running median and percentiles for the upper limits. Histograms on the right show the cumulative distribution of the color and upper limits. Shaded grey areas depict the typical color of a Type I QSO with high obscuration at different redshifts. The median color of F1800W detections, F770W$-$F1800W$=1.4^{+0.4}_{-0.6}$~mag, is much bluer than a dust-obscured AGN, with only two LRDs showing such IR red colors. Similarly, about 75\% of F1800W non-detections have F770W$-$F1800W$<2.7$~mag, bluer than an obscured QSO at $z\gtrsim4$.}
\label{fig:MIRI_color}
\end{figure}

As noted in \citet{williams24} and \citet{pg24a}, the small detection fraction in F1800W appears to be driven by a combination of the shallow depth of the F1800W data and the relatively flat rest-frame IR SED of the LRDs (see also \citealt{akins24}).  Figure~\ref{fig:MIRI_color} shows the F770W$-$F1800W colors for the MIRI-detected LRDs as a function of F444W. Non-detections in F1800W are indicated as upper limits computed using the $5\sigma$ limiting magnitude. Overall, the median color of the F1800W detections, F770W$-$F1800W$=1.4$~mag, agrees well with previous works (e.g., \citealt{leung24}) and confirms that the IR emission in most LRDs differs substantially from the rising power law of dust-obscured QSOs, which have redder colors (F770W$-$F1800W$>2$~mag) at $z\gtrsim4$. However, we identify two LRDs, COS-50808 and COS-40053, with red IR colors indicative of strong hot dust emission. Notably, these LRDs are not among the most obscured in the optical, but instead exhibit some of the bluest UV-to-NIR colors. In contrast, the LRD with F770W$-$F1800W$\sim2$~mag, UDS-47101, is the best candidate for an obscured AGN. We further discuss these sources in the next section. There are also 2 LRDs in the opposite side of the spectrum, with blue IR colors (F770W$-$F1800W$<0.5$~mag), indicating a remarkably flat or even declining SED in the rest-frame NIR (UDS-22702 and UDS-40579; the latter is discussed in \S~\ref{s:lowz_LRDs}). In the next section, we use the full SEDs and redshifts to estimate rest-frame colors and provide more meaningful constraints on the shape of their near-to-mid IR emission, showing how they compare to galaxy- or AGN-dominated models.

Figure~\ref{fig:selection_MIRI} also shows the running median (blue line) and cumulative color distribution of the F1800W non-detections (right panel) indicating that nearly half the sample has upper limits bluer than a typical obscured QSO (F770W$-$F1800W$\lesssim2$~mag). The brightest non-detections at F444W$\lesssim25$~mag (approximately 25\% of the sample) place strict upper limits on the F770W$-$F1800W colors, which are bluer than the median of the F1800W detections. Conversely, the faintest non-detections at F444W$\gtrsim25$~mag provide only loose constraints on the F770W$-$F1800W color, leaving the possibility of hot dust emission mostly unconstrained.

\begin{figure*}%[htp!]%[ht!]
\centering
\includegraphics[width=18cm,angle=0]{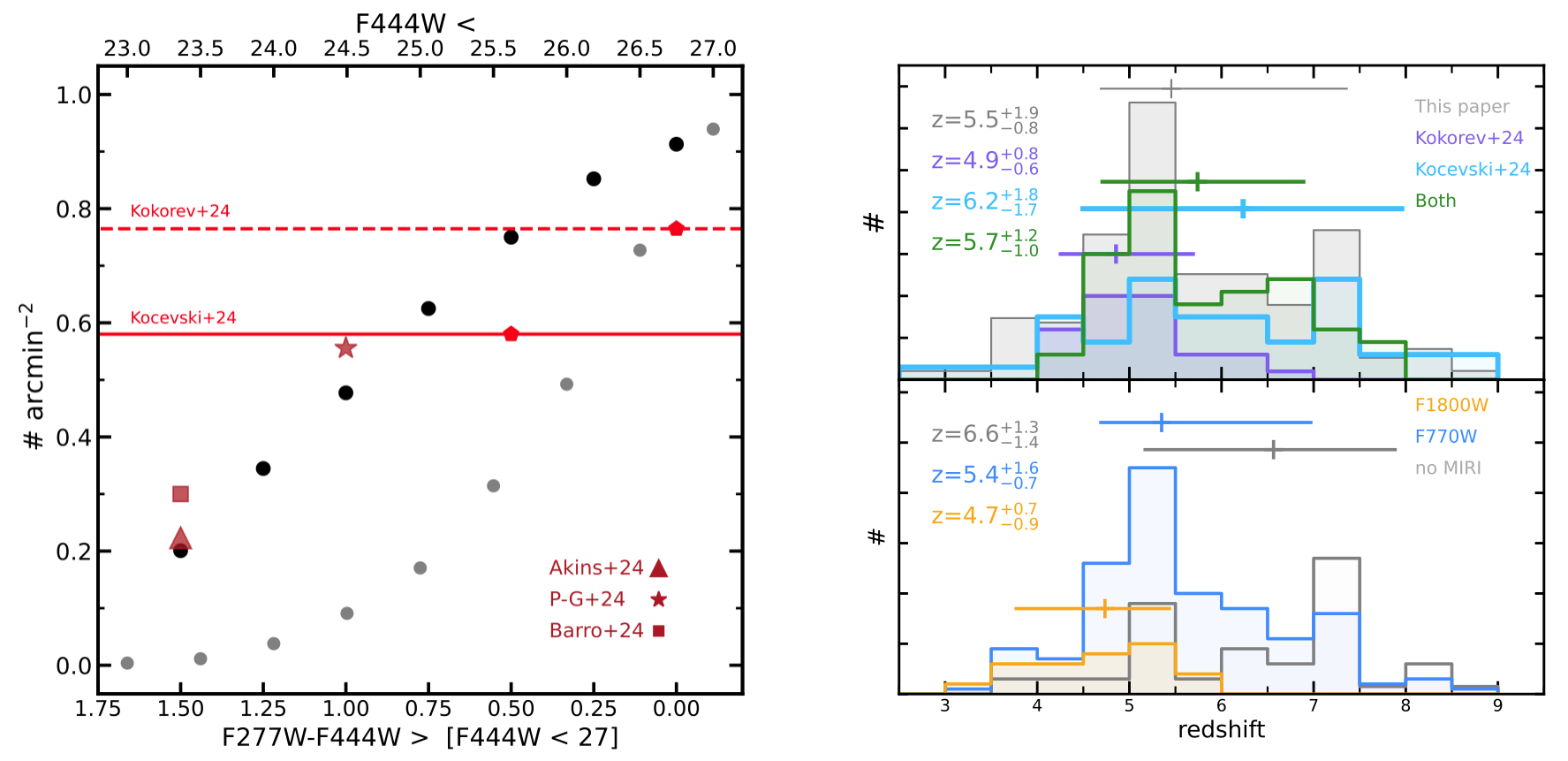}
\caption{{\it Left:} Cumulative number density of LRDs as a function of F277W$-$F444W color (black; bottom axis) and F444W magnitude (grey; top axis). The density increases steeply with bluer colors, rising by up to $\sim$5$\times$ from F277W$-$F444W$=$1.5 to 0. Our selection agrees with previous samples (red markers) within similar thresholds but identifies more LRDs overall, with a total density of 0.93~arcmin$^{-2}$ down to F444W$=27$~mag. {\it Top-right:} Redshift distribution of the full LRD sample (grey) compared to subsets identified in KOV24 (blue), KOK24 (purple), or both (green). The median redshift, $z=$5.5, is lower than in KOI24 ($z\sim6.2$), likely due to $z\lesssim5$ LRDs with blue colors F277W$-$F444W$\lesssim1$~mag caused by \OIII\ flux contamination, found only in KOV24 (purple). {\it Bottom-right:} Redshift distributions by MIRI properties. F1800W detections (orange) cluster at $z\lesssim5$, F770W detections (blue) follow the overall sample, while non-detections (grey) are predominantly at higher redshifts, $z=6.6$.} 
\label{fig:density}
\end{figure*}

\subsection{LRD Sample: Number densities}

Figure~\ref{fig:density} shows the cumulative number density of LRDs in our sample as a function of F277W$-$F444W color (with F444W$<27$~mag; black circles, bottom axis) and F444W magnitude (any F277W$-$F444W color; grey circles, top axis), compared to other samples from the literature. We find a total density of 0.93 arcmin$^{-2}$ to F444W$=27$~mag, which could exceed 1 arcmin$^{-2}$ by extending the depth by 0.5 to 1~mag. The number density from
\citet{akins24}, selected with a similar limit of F444W$<27.2$~mag but over an area nearly seven times larger, agrees well with our values at F277W$-$F444W$>1.5$~mag, suggesting this criterion offers a robust and low-contamination selection. The densities from \citet{barro24} and \citet{pg24a}, identified with color thresholds of F277W$-$F444W$>1$ and 1.5~mag over smaller areas, exhibit a similar color trend to our sample, with only slightly higher values, by factors of $\times$1.3 and $\times$1.1, respectively, when applying the same magnitude limit of F444W$<27$~mag. The comparison to KOV24 and KOI24 (red dashed and solid lines) is more complex because their selections are not solely based on the F277W$-$F444W color. However, since Figure~\ref{fig:selection_othersamples} shows that $\sim$90\% of their LRDs have F277W$-$F444W$>0$~mag and F277W$-$F444W$>0.5$~mag, we compare to our sample at those approximate limits (red hexagons). Even under these conditions, their densities remain lower than ours by a factor of $\times$1.3, despite their deeper magnitude limits (F444W$\lesssim28$~mag at SNR$_{\rm F444W}<12$ and F444W$<27.7$~mag). This is consistent with the results in the previous section showing that the overlap between their samples is only about 50\%.

The differences in the number of LRDs selected by different criteria have direct implications for the luminosity and stellar mass functions of LRDs derived in prior works, which estimate densities on the order of 10$^{-5}$~Mpc$^{-3}$ (e.g., \citealt{greene23}; \citealt{akins24}) based on a single selection method. The new comprehensive selection reveals nearly $\times$1.7 more LRDs, indicating that previous studies may have substantially underestimated the true volume densities of these sources.

\section{Properties of the LRDs}
\label{s:properties}
\begin{figure*}%[htp!]%[ht!]
\centering
\includegraphics[width=18cm,angle=0]{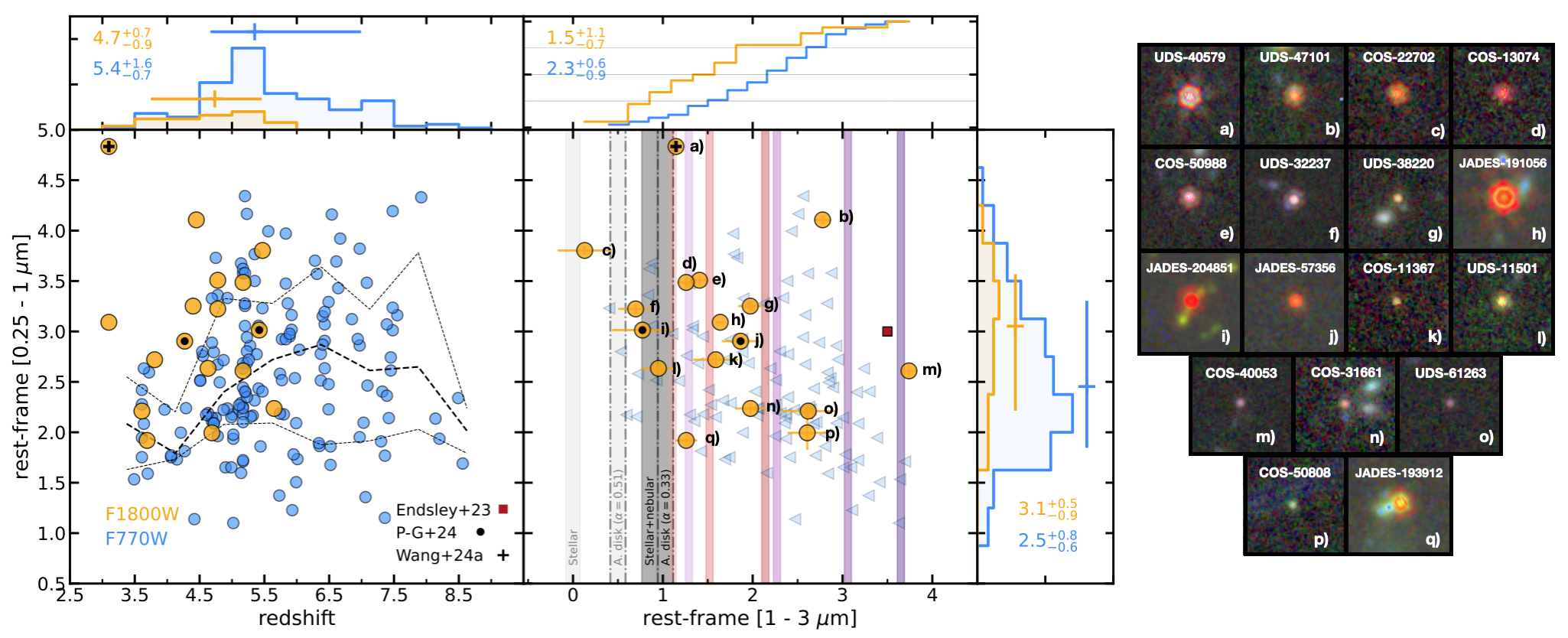}
\caption{{\it Left:} UV-to-NIR color vs. redshift for LRDs detected in F1800W (orange) and F770W (blue). Black and red markers indicate LRDs in common with previous works. LRDs span a broad range of $\sim3$~mag in UV-to-NIR colors, with medians of \uvnir$=2.5$~mag (F770W) and 3.0 (F1800W). The dashed line shows the running median and percentiles, highlighting a trend toward redder colors at $z\gtrsim5$, driven by a lack of very red LRDs at lower redshifts. {\it Center:} Center: UV-to-NIR vs. near-to-mid IR colors for F1800W-detected LRDs (orange) and upper limits for undetected ones (blue triangles). Only LRDs at $z<6$ are shown. The vertical lines represent \nearmidir colors for galaxy and AGN SED models from Figure~\ref{fig:midIR_templates} with the same color code. The \nearmidir colors have a wide scatter and a median of \nearmidir$=1.5$~mag. Only $\sim20$\% of F1800W detections have negligible dust emission, while the bulk ($\sim50$\%) show dust emissions of up to f\,$^{\rm dust}_{3 \mu m}=0.7$ and 0.8 from the ISM or AGN torus. F1800W detections show a negative correlation, suggesting that bluer UV-to-NIR LRDs have redder IR colors (indicating more dust emission), resembling type I QSOs. {\it Right:} NIRCam 2''$\times$2'' color composite (F150W+F277W+F444W) cutouts of F1800W-detected LRDs, sorted by UV-to-NIR colors from red to blue.}
\label{fig:ircolors}
\end{figure*}

\subsection{Redshift distributions}
\label{s:redshifts}

The right panel of Figure~\ref{fig:density} shows the redshift histograms for the full LRD sample and the subsets discussed earlier. The distribution spans $3<z<9$, with primary and secondary peaks at $z=5$ and $z=7$, and a median of $z=5.5$. While qualitatively similar to KOI24, our redshift distribution has a lower median redshift than theirs, $z\sim6.4$. This difference likely reflects the rapid decline in our selection efficiency at $z\gtrsim8$ (see \S~\ref{s:criteria}) and the stronger $z=5$ peak in our sample. This peak is driven by additional LRDs present in KOV24 but absent in KOI24 (purple histogram, peaking at $z=4.9$), a subset characterized by bluer F277W$-$F444W$\lesssim1$~mag colors due to strong \OIII\ emission boosting the flux in F277W.

The bottom panel histograms, divided into MIRI subsets, show that F770W detections, which make up approximately 77\% of the sample, closely follow the distribution of the overall population. The only exception is the small fraction of F444W-faint LRDs not detected in MIRI, predominantly found at higher redshifts. In contrast, F1800W detections are concentrated at much lower redshifts, with a median of $z=4.7$. The brighter median magnitudes and lower redshifts suggest that detectability in F1800W is at least partly driven by a mid-IR luminosity threshold. However, as discussed in \S~\ref{s:luminosities}, intrinsic variations in the near-to-mid IR colors of LRDs also affect the detection fraction.

\subsection{Rest-frame UV-to-NIR colors}
\label{s:restcolors}

The left panel of Figure~\ref{fig:ircolors} shows the rest-frame UV-to-NIR color \uvnir vs. redshift for the MIRI-detected LRDs. We focus on this subset because F770W is critical for probing the rest-frame 1~$\mu$m up to $z\sim8$. The distribution spans a broad color range from \uvnir$=1$ to 5~mag, with a median of 2.5~mag, and is slightly redder for the F1800W detections, with a median of 3.1~mag. This large scatter is consistent with the results from \citet{pg24a} for the small SMILES sample, confirming that LRDs exhibit a wide diversity of intrinsic colors within the same characteristic V-shaped SED. At the extremes, $\sim10$\% of LRDs have colors bluer than \uvnir$\lesssim1.6$~mag, i.e., their SEDs are between the bluest LRD model from \citet{pg24a} and the type 1 QSO templates in Figure~\ref{fig:selection_criteria}, suggesting that they may resemble traditional, low-obscuration AGNs. Conversely, $\sim3$\% of LRDs exhibit extremely red colors (\uvnir$>4$~mag), falling between the reddest LRD model from \citet{pg24a} and the LRD from \citet{wang24_lrd}, the reddest in our sample. The running median of the distribution (black dashed line) indicates a trend toward redder colors with increasing redshift, shifting from \uvnir$\sim2$ to 2.7~mag at $z\gtrsim4.5$. This trend is driven by the absence of very red LRDs (\uvnir$>3$~mag) at $z\lesssim4.5$, with only a few F1800W detections being exceptions. The median color and scatter remain relatively constant, with no significant differences in the colors of LRDs between $z=5$ and $z=7$.

\subsection{Rest-frame near-to-mid IR colors}
\label{s:restcolors2}

In this section, we examine the near-to-mid IR colors of MIRI-detected LRDs and compare them to predictions from galaxy and AGN-dominated models. The \nearmidir color probes a key spectral region where dust emission—either hot or warm—could contribute significantly to the spectrum, potentially dominating over the stellar or AGN continuum that dominates the optical-to-NIR SED. Redder mid-IR colors would indicate a stronger, and possibly dominant, contribution from dust emission. To ensure accurate rest-frame colors, we include only LRDs up to $z=6$, beyond which the rest-frame 3$\mu$m begins to shift out of the F1800W filter. For LRDs undetected in F1800W, we use the 5$\sigma$ upper limit to model the SED and calculate the colors. In the following, we present the different IR SED models (\S~\ref{s:IRmodels}) and compare the \uvnir vs. \nearmidir colors of the LRDs to the model predictions (\S~\ref{s:colorcompare}). Appendix~\ref{s:f1800w_seds} shows the best-fit SEDs and rest-frame colors for the F1800W detections.

\begin{figure}
\includegraphics[width=9cm,angle=0]{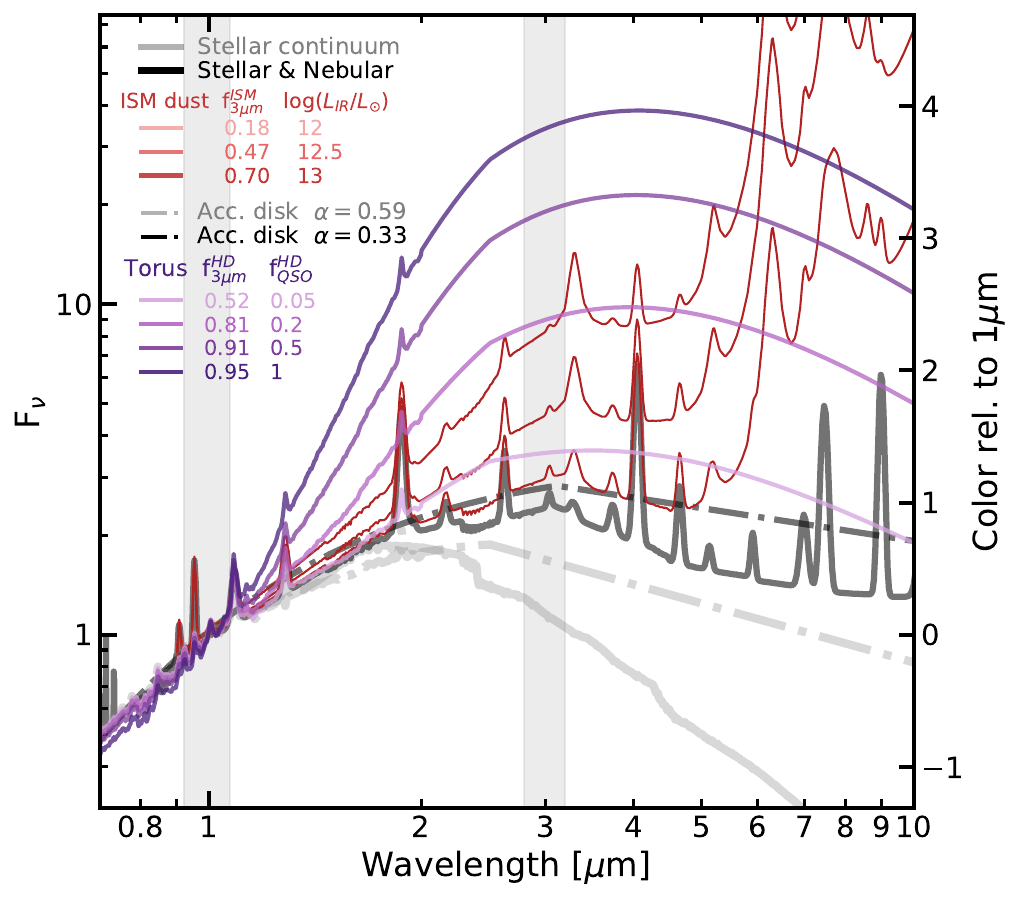}
\caption{Stellar and AGN-dominated SED models illustrating the range of near-to-mid IR colors for LRDs. The solid lines represent models where the continuum is dominated by stars (light) or stars plus nebular emission (dark). The dashed lines correspond to models dominated by an accretion disk with power-law slopes of $\alpha=0.59$ and 0.33 (light and dark). The magenta and purple lines show models with dust emission from the galaxy ISM or AGN torus, respectively. Darker colors indicate increasing ISM-dust or torus-dust emissions relative to the stellar or accretion disk continua at 3$\mu$m.}
\label{fig:midIR_templates}
\end{figure}

\subsubsection{IR SED models}
\label{s:IRmodels}

Figure~\ref{fig:midIR_templates} presents several galaxy and AGN-dominated models proposed in recent studies (e.g., \citealt{pg24a}; \citealt{akins24}; \citealt{leung24}) to describe the IR emission of LRDs. We compare these models against the \nearmidir colors of the LRDs in the following section. The models are divided into two groups: those with dust emission and those without. 

{\bf Without dust emission:} In these models (light and dark grey), the continuum emission is either from stars (solid line) or an AGN accretion disk (dashed line), providing a baseline for tracing additional dust emission. When stellar emission dominates, the SED shows a characteristic 1.6~$\mu$m peak and declines at longer wavelengths, yielding blue colors (\nearmidir$=0$~mag) that can increase up to \nearmidir$=1$~mag if nebular emission is also present. If the continuum is dominated by an accretion disk with significant attenuation (A(V)$=4$), modeled as a power-law from $\lambda>5000$~\AA\ (f$_{\nu}\propto\lambda^{-\alpha}$), the colors range from \nearmidir$\sim0.5$ to 0.9~mag (light to dark grey), depending on the slope values between $\alpha=0.593$ and 0.33, as in \citet{temple21} and \citet{hernancaballero16}. Note that the default QSO model of \citet{temple21} includes a stellar component contributing to the NIR flux; here, we only show the accretion disk SED.

{\bf With dust emission:} Dust emission from the galaxy ISM can increase the \nearmidir colors relative to the stellar continuum, depending on the dust temperature and mass. Cold dust (T$\sim30$~K) peaking at $\lambda\sim100~\mu$m, typical in dusty SFGs, is inconsistent with non-detections in deep ALMA 1.1~mm stacks of LRDs (e.g., \citealt{labbe24}; \citealt{williams24}; \citealt{akins24}), but warmer dust (T$\sim50$~K) peaking at $\lambda\sim60~\mu$m could explain both ALMA and Herschel non-detections (\citealt{pg24a}; \citealt{akins24}; \citealt{casey24}). These high temperatures could arise from the compact sizes and intense radiation fields of LRDs, as seen in the starburst models of \citet{siebenmorgen07}. The magenta lines, with progressively darker colors, show models with increasing IR luminosities from $\log(L_{\rm IR}/L_{\odot})=12$ to 13. The fraction of ISM dust contributing to the SED at $\lambda=3~\mu$m in these models ranges from f\,$^{\rm ISM}_{3\mu m} \sim0.2$ to 0.7, leading to increasingly red colors, from \nearmidir$=1$ to 2.1~mag.

Similarly, dust emission from an AGN torus can also increase the \nearmidir colors relative to the accretion disk continuum, depending on the dust temperature and distribution in the central few parsecs. While the relatively flat IR SED of LRDs
(\citealt{williams24}; \citealt{pg24a}; \citealt{leung24}) suggest that hot dust (T$\sim$1200K) is not dominant, indicating a hot-dust deficiency relative to typical QSOs (e.g., \citealt{lyu17}), smaller hot-dust fractions or lower dust temperatures could still be consistent with the MIRI data. For example, following \citet{leung24}, the purple lines in Figure~\ref{fig:midIR_templates} show increasing hot-dust fractions from a modified black body (MBB) with $T=1240$~K. The hot-dust fraction relative to the accretion disk continuum at $\lambda=3~\mu$m ranges from f\,$^{\rm HD}_{3\mu m}\sim0.5$ to 0.95. Note that \citet{leung24}, quotes instead hot-dust fractions relative to the median QSO model of \citet{temple21} where the hot-dust emission is $\times$4 the flux of the accretion disk at $\lambda=2~\mu$m. Hence, f\,$^{\rm HD}_{\rm QSO}=1$ represents a typical dust-obscured QSO, hot-dust dominated (f\,$^{\rm HD}_{3\mu m}=0.95$), with a very red color \nearmidir$=3.6$~mag. More complex models, such as combinations of MBBs at lower temperatures or radiative transfer models like CLUMPY \citep{nenkova08} and SKIRTOR \citep{stalevski12}, can also reproduce the IR SEDs of LRDs (\citealt{pg24a}; \citealt{wang24_lrd}; \citealt{akins24})

The similarities between the ISM and torus emission models in Figure~\ref{fig:midIR_templates} suggest that the 1 to 3~$\mu$m spectral range is too narrow to effectively resolve the SED degeneracies and discriminate between galaxy- or AGN-dominated scenarios for LRDs. This underscores the need for additional MIRI photometry, particularly between F770W and F1800W, as in the SMILES survey \citep{albers24}, MEGA (Kirkpatrick, in prep.), or the upcoming MEOW (PI: Leung) surveys. Such data would enable more detailed SED modeling of the near-to-mid IR range.

\begin{figure*}%[htp!]%[ht!]
\centering
\includegraphics[width=18.5cm,angle=0]{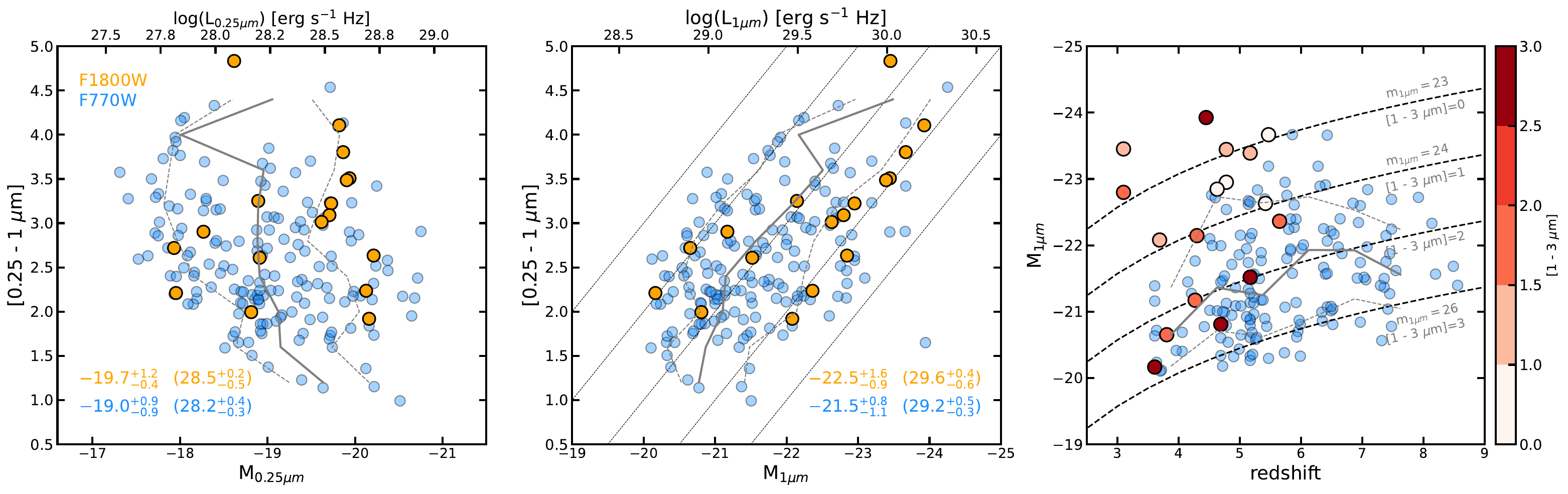}
\caption{UV and NIR absolute magnitudes (M$_{0.25 \mu m}$ and M$_{1 \mu m}$) vs. color and redshift for F1800W detections (orange) and non-detections (blue). {\it Left:} Running medians (grey lines) show a constant median M$_{0.25 \mu m}=-19$~mag with $\sim$2~mag scatter, consistent with previous works.  {\it Central:} NIR luminosity has similar scatter, with a median of M$_{1.0 \mu m}=-21.5$~mag and weak correlation with UV-to-NIR color.  {\it Right:} NIR luminosity vs. redshift color-coded by \nearmidir for F1800W detections. Dashed lines trace absolute magnitude limits for constant observed magnitudes at 1~$\mu$m ($\sim$~F770W). F1800W detections are largely limited to bright LRDs, M$_{1 \mu m}\lesssim-22$~mag, although $\sim$30\% are detected at lower luminosities due to their red colors, \nearmidir$=2.2$~mag.}
\label{fig:luminosity_color}
\end{figure*}

\subsubsection{UV-to-NIR vs. near-to-mid IR colors}
\label{s:colorcompare}

The central panel of Figure~\ref{fig:ircolors} shows the rest-frame near-to-mid IR ([1$-$3 $\mu$m]) colors and upper limits versus the UV-to-NIR ([0.25$-$1 $\mu$m]) colors for the MIRI-detected LRDs, compared to the IR models from Figure~\ref{fig:midIR_templates}. The colors of the models are represented as vertical lines because the UV emission at 0.25~$\mu$m is not necessarily correlated with the IR emission; it can arise from a secondary (unobscured or patchy) component with an arbitrary UV luminosity.

The F1800W-detected LRDs have a median color of \nearmidir$=1.5$~mag, which is significantly bluer than the typical obscured QSO by $\sim$2~mag, as noted in previous studies (\citealt{williams24}; \citealt{pg24a}). However, the color distribution spans a broad range of $\sim$2.5~mag, indicating that LRDs exhibit a wide diversity of \nearmidir colors and, consequently, dust emission properties. About $\sim20$\% of the F1800W detections have blue \nearmidir$\lesssim1$~mag colors, suggesting negligible dust emission and nearly flat SEDs, similar to the MIRI stacks from \citet{williams24}, \citet{akins24} (\nearmidir$\sim0.9$~mag), or \citet{casey24} (\nearmidir$=-0.3$~mag). Approximately 50\% of the population has colors between $1<$\nearmidir$\lesssim2.25$~mag, indicating more substantial dust emission, contributing up to 70\% – 80\% of the flux at 3~$\mu$m, either from ISM dust or a hot-dust torus. However, this is still small relative to a dust-obscured QSO (f\,$^{\rm HD}_{\rm QSO}\lesssim20$\%), in agreement with the findings of \citet{leung24}. Among the reddest 20\%, only one object (UDS-40053, {\it m)}) displays exceedingly red colors (also in F770W$-$F1800W), nearly 1~mag redder than the others, resembling a dust-obscured QSO. Interestingly, its UV-to-NIR color is bluer than the median (see discussion below). The LRD in the top-right corner (UDS-47101, {\it b)}) is also notable for being among the reddest LRDs in UV-to-NIR color, second only to the most extreme LRD in \citet{wang24_lrd} (UDS-40579, {\it a)}). However, it exhibits a red UV slope, making its overall SED resemble a dust-obscured AGN at lower redshifts (e.g., the IRAC power-law AGNs in \citealt{donley12,donley18}) rather than a typical LRD. For comparison, we also show the location of a radio-loud QSO at $z\sim7$ from \citet{endsley23}, which hosts an obscured AGN but is not as red as the F1800W-detected LRDs in the rest-frame optical.

Attending to the overall trend, it appears that the \nearmidir color of the F1800W detections follows a negative correlation with \uvnir, i.e., LRDs with bluer UV-to-NIR colors exhibit stronger dust emission. While this may seem counterintuitive, it is consistent with the framework discussed in \S~\ref{s:sample} (e.g., Figure~\ref{fig:selection_criteria}), where LRDs with blue UV-to-NIR colors might resemble type I QSOs, which are IR-bright due to hot dust dominating the emission at $\lambda>1\mu$m. In this scenario, LRDs similar to unobscured QSOs would populate the bottom-right region of the color-color diagram, while obscured, power-law QSOs would populate the top-right (e.g., as UDS-47101).

While only 7\% of the LRDs are detected in F1800W, the large number of non-detections still provide valuable constraints on the \nearmidir colors. The blue arrows and cumulative histogram in the central panel of Figure~\ref{fig:ircolors} show the distribution of the \nearmidir upper limits, with a median of \nearmidir$<2.3$~mag. This implies that about half of the LRDs have dust emissions weaker than those in the bulk of the F1800W detections. The most restrictive constraints come from the brightest LRDs (F770W$\lesssim24$~mag), which make up the bluest top $\sim10$\%, with \nearmidir$\lesssim1$~mag, consistent with negligible dust emission. The fainter top quartiles are the least restrictive, although they still confirm that the dust emissions in these LRDs are much lower than those in hot-dust-dominated QSOs (f\,$^{\rm HD}_{\rm QSO}\lesssim$0.5).

\subsection{UV and NIR luminosities}
\label{s:luminosities}

The left and central panels of Figure~\ref{fig:luminosity_color} show the \uvnir color vs. absolute magnitudes (luminosities) at 0.25 and 1~$\mu$m. The UV absolute magnitude has a median of M$_{0.25 \mu m} = -19$~mag and a scatter of $\sim2$~mag, consistent with typical UV luminosity functions from previous works (\citealt{greene23}; \citealt{kocevski24}). The NIR absolute magnitude has a median of M$_{1 \mu m} = -21.5$ and similar scatter. The F1800W detections are typically brighter by 0.7 and 1~mag in the UV and NIR, respectively. The running medians (grey lines) suggest a weak correlation between UV-to-NIR color and NIR luminosity, with redder \uvnir colors possibly driven by larger NIR luminosities. This trend is more pronounced in the brightest F1800W detections, which show redder colors (\uvnir$=2$ to 4~mag) with increasing magnitudes (M$_{1 \mu m} = -22$ to $-24$~mag). In a multi-component SED model for LRDs, this could indicate that the UV-to-NIR color traces increasing AGN luminosity relative to the unobscured galaxy host.

The right panel of  Figure~\ref{fig:luminosity_color} shows the NIR absolute magnitude versus redshift for the F1800W detections, color-coded by their \nearmidir colors. Dashed lines indicate the absolute magnitudes corresponding to constant observed magnitudes at rest-frame 1$\mu$m (probed by F770W at $z\gtrsim5$), ranging from 23 to 26~mag. These lines approximate \nearmidir upper limits, assuming F1800W probes rest-frame 3$\mu$m with a 5$\sigma$ detection limit of F1800W$=23$~mag. In this grid, F1800W non-detections (blue circles) have \nearmidir colors bluer than the dashed line, while detections are redder (e.g., those above the M$_{1 \mu m}=24$~mag line must have \nearmidir$<1$~mag; otherwise, they would be detected). These lines reflect the same trends as the middle panel of Figure~\ref{fig:ircolors}  but emphasize the dependence on M$_{1 \mu m}$. The top 70\% of F1800W detections, at M$_{1 \mu m}\lesssim-22$~mag, represent the most luminous LRDs across redshifts and exhibit a bluer median \nearmidir$=1.4$~mag. This suggests that their F1800W detection is driven by high NIR luminosities, despite relatively modest mid-IR emission. Conversely, the bottom 30\% have redder \nearmidir$=2.2$~mag colors, indicating they are detected primarily due to stronger mid-IR luminosity, despite being fainter in the NIR.

\section{Search for dust emission signatures in low-z LRDs}
\label{s:lowz_LRDs}

The similarity in \nearmidir colors of the different LRD SED models makes it challenging to determine whether the dust emission originates from the galaxy's ISM or an AGN torus. While improving the IR SED coverage with additional MIRI bands (e.g., SMILES or MEGA surveys) could help distinguish between dust models, the high median redshift of LRDs ($z\sim6$) places the peak of their IR emission beyond MIRI's wavelength range ($\lambda_{\rm rest}\sim$30–60$\mu$m; \citealt{pg24a}; \citealt{akins24}; \citealt{casey24}). However, as the stellar or accretion disk continuum declines rapidly into the mid-IR, dust emission, even if modest, must dominate at $\lambda_{\rm rest}\gtrsim3$. Targeting lower-redshift LRDs offers a promising path forward, as they are more likely to be brighter, enabling detections in MIRI's longer-wavelength bands and providing tighter constraints from far-IR to radio data.

The best candidates for this analysis are two LRDs with spectroscopic redshifts at $z\sim3.1$: UDS-40579 and JADES-191056 ({\it a)} and {\it h)} in Figure~\ref{fig:ircolors}). These sources are over a magnitude brighter (F444W$\lesssim22.5$~mag) than other LRDs at $z<4$. UDS-40579 was studied in detail by \citet{wang24_lrd}, and our SED modeling yields consistent results. This re-analysis allows a comparison
with JADES-191056 and stidues the ISM dust emission model not covered previously. {\bf UDS-40579} exhibits bright MIRI detections in F770W and F1800W and NIRSpec spectroscopy showing strong, broad AGN emission lines. Its relatively blue IR color (\nearmidir$=1.1$~mag) suggests minimal dust emission, yet its extremely red UV-to-NIR color (\uvnir$=4.9$~mag) makes it very different from typical LRDs, which have a median \uvnir$=2.5$~mag. {\bf JADES-191056} is similarly bright and has optical spectroscopy from the GOODS-S MUSE-Wide survey (showing Lyman-$\alpha$ emission; \citealt{kerutt22}). However, its UV and IR colors (\uvnir$=3.1$~mag, \nearmidir$=1.6$~mag) are more representative of the LRD population. The extensive MIRI coverage from the SMILES survey provides photometry in seven bands, up to F2100W. Furthermore, we also find a significant  ($\sim9\sigma$) detection in Spitzer/MIPS24 imaging (GOODS/FIDEL; \citealt{dickinson03}) and weaker detections (SNR$\sim$4) in Herschel/PACS100 and PACS160 mosaics (PEP; \citealt{berta11}; \citealt{lutz11}). While it appears in HerMES SPIRE 250 and 350 catalogs \citep{magnelli13}, the fluxes have very low SNR. We use upper limits instead for all SPIRE bands. Additionally, it is undetected in the deep GOODS-ALMA 1.1~mm map, with a limiting 5$\sigma$ sensitivity of $\sim500~\mu$J beam$^{-1}$ (\citealt{franco18}; \citealt{gomezguijarro22a}). 

The lower spatial resolution of Spitzer and Herschel (FWHM$\sim$5'', 7'', and 11'') compared to even the reddest MIRI band (FWHM$=$0.67'' in F2100W), complicates the flux measurement for JADES-191056, which has a nearby companion at $\sim$2.4'' with a similar redshift. While resolved and detected in all MIRI bands, the companion is blended with the LRD in MIPS24, making the photometry more challenging. In Appendix~\ref{s:mips_flux}, we outline the photometric procedure used to obtain precise fluxes for the LRD and its companion in MIPS24, leveraging the higher spatial resolution of MIRI imaging. For PACS100 and PACS160, the combination of lower spatial resolution and lower SNR prevents an accurate photometric deconvolution. Instead, we use total fluxes measured within circular apertures, as described in \citet[][see also \citealt{2010A&A...518L..15P} for a discussion on Herschel photometry with position priors]{barro19}, and assign a larger uncertainty to account for the flux contribution of the companion, estimated at $\sim$30\% from the MIPS24 analysis.

In the following sections, we examine the UV and optical morphologies of these two LRDs and perform a detailed SED analysis using AGN- and galaxy-based models to investigate the origins of their continuum emission in different spectral ranges.

\begin{figure*}%[htp!]%[ht!]
\centering
\includegraphics[width=18cm,angle=0]{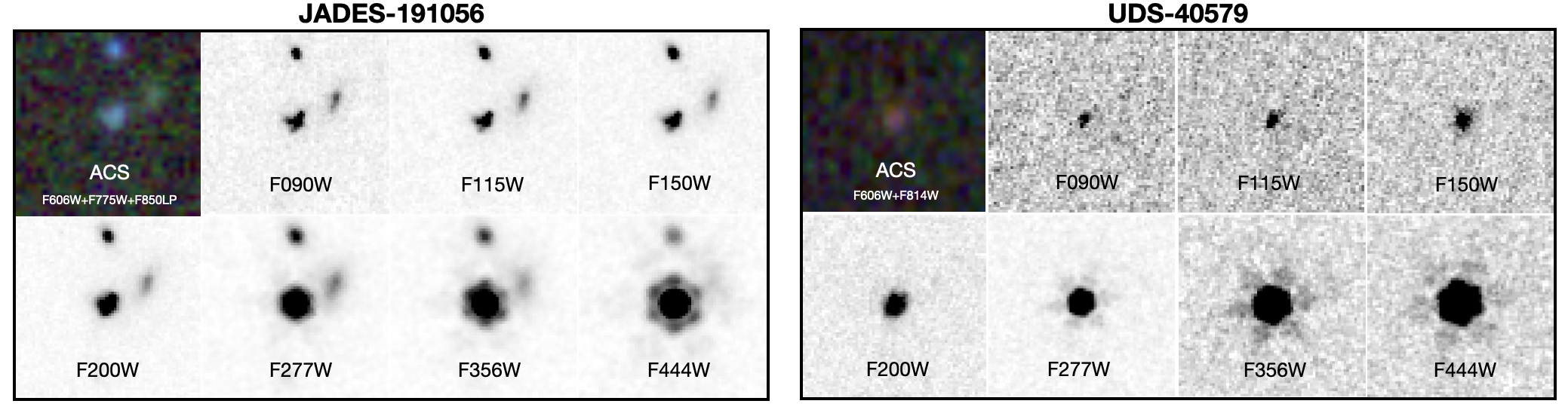}
\caption{2''$\times$2'' cutouts of the two F1800W-detected LRDs at z$=$3.1 in different NIRCam bands. {\it Left:} JADES-191056 shows resolved structure in the short-wavelength NIRCam bands with a low S\'ersic index ($n=0.58$) and r$_{e}=0.62$~kpc in F090W. The source is slightly elongated, possibly with a secondary component and a few close companions. In long-wavelength bands (F200W and onward), which probe the rest-frame optical, it appears point-like. The HST/ACS color composite (F606W+F775W+F850LP) confirms its blue rest-frame UV color and shows a similar morphology. {\it Right:} UDS-40579 appears compact and unresolved in all NIRCam bands with no clear features. The HST/ACS color composite is fainter, redder, and indicates a steeper UV slope than JADES-191056.} 
\label{fig:LRD_galfit}
\end{figure*}

\subsection{Morphologies}

The low redshift of these LRDs enables the study of their rest-frame UV morphology using high spatial resolution imaging from HST and JWST at observed wavelengths between $0.6<\lambda<1~\mu$m (FWHM$\sim0.03$'' to 0.08''). To analyze their structure, we compute the effective radius (r$_e$) along the major axis and \citet{sersic} indices (n) in 7 bands from F090W to F444W using GALFIT v3.0.5 \citep{galfit}, following the procedure in \citet{pgp23a} for centering, masking of nearby sources, and estimating the ERR array (see McGrath in prep. for additional details). Empirical PSFs were created using stars from the JADES and UDS fields. Figure~\ref{fig:LRD_galfit} shows multi-band cutouts for the two LRDs in several JWST/NIRCam bands, along with an HST color composite.

As noted above, UDS-40579 is the reddest LRD in the sample in UV-to-NIR color, making it faint in both HST/ACS and JWST/NIRCam SW bands (F150W$\sim$27.5~mag). It appears faint, red, and compact across all bands, with no clear signs of resolved structure. The S\'ersic fits return the minimum possible values, indicating a point-like, unresolved object, consistent with the findings of \citet{wang24_lrd} using \textsc{pysersic} \citep{pysersic}. In contrast, JADES-191056 is more than one magnitude brighter at observed wavelengths $\lambda<1~\mu$m, has bluer colors, and exhibits clear structural features. The HST/ACS and JWST bands up to F150W show a resolved, elongated morphology towards the NE, with a possible smaller secondary structure crossing its center in the perpendicular direction. The transition to point-like morphology at F200W coincides with the change in the SED from flat UV to steep, red optical. At F277W, this is likely influenced by contamination from a strong \OIII\ emission line. The S\'ersic index reflects this trend, with small, disk-like values in F090W and F115W (n$=0.57$, 0.58) increasing in the redder bands (n$=0.85$ to 2 in F150W and F200W), indicating a more compact region dominating the optical emission. The effective radii in F090W and F115W are very similar (r$_{e}=0.09$'' and 0.08''), corresponding to a physical, non-circularized radius of 0.62~kpc, which is consistent with typical star-forming LBGs at $z\sim3$ with M$_{0.25 \mu m}=-19.7$~mag (e.g., \citealt{shibuya15}).

This is consistent with previous findings, which suggest that LRDs are slightly more resolved in the rest-frame UV than in the optical (e.g., \citealt{baggen23}). The resolved UV morphology supports models where LRDs are a mix of a blue, low-mass galaxy hosting an obscured, compact component—either an AGN or a strong star formation burst. In the latter case, the star formation burst contains most of the total stellar mass. This possibility, however, contrasts with other massive compact star-forming galaxies typically observed at lower redshifts ($2<z<3$; \citealt{barro13, nelson14}). These galaxies also host dusty nuclear starbursts visible in ALMA observations at sub-kpc scales (e.g., \citealt{barro16, tadaki17a}), but they exhibit galaxy-wide obscuration (i.e., the entire galaxy appears red), rather than a red burst hidden within a blue host.

\begin{figure*}%[htp!]%[ht!]
\centering
\includegraphics[width=18cm,angle=0]{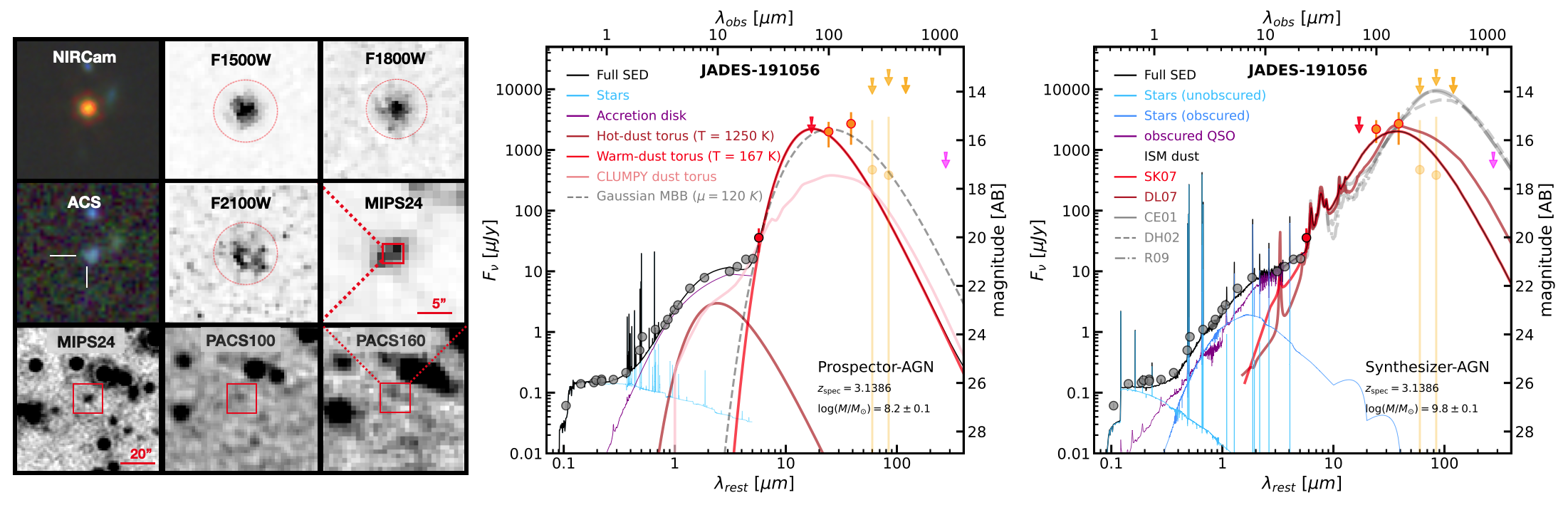}\\
\includegraphics[width=18cm,angle=0]{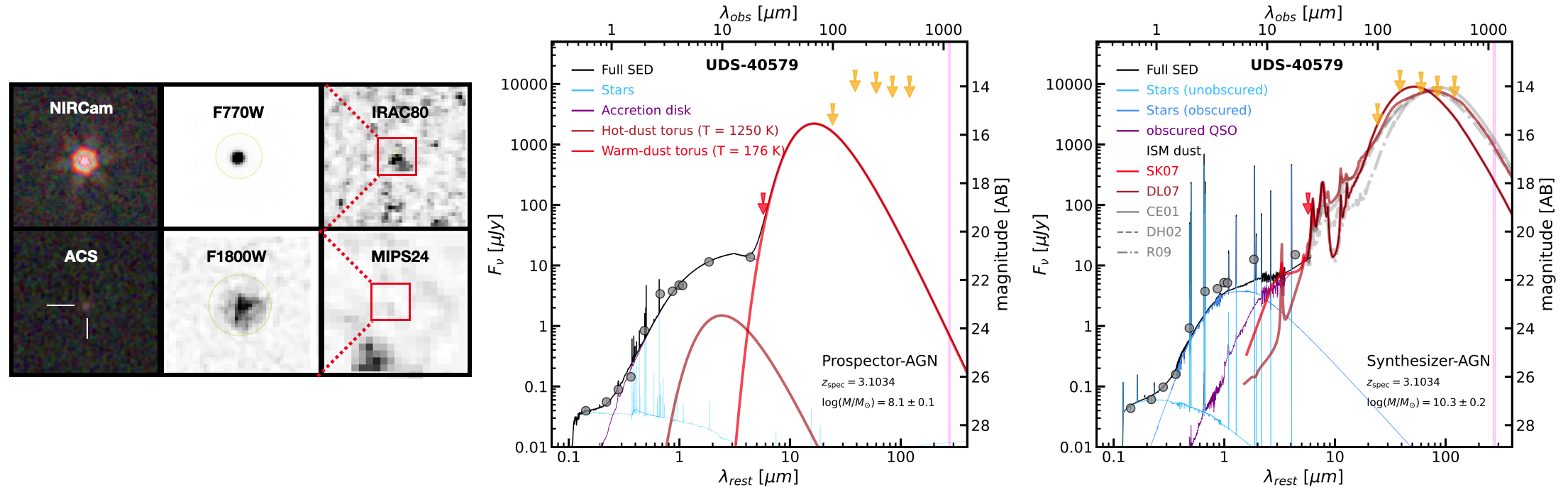}
\caption{5''$\times$5'' cutouts and best-fit SEDs for the two LRDs at $z\sim3.1$, using galaxy- and AGN-dominated models. 
{\it Top:} JADES-191056. {\it Left:}  The top rows show the blue rest-frame UV (ACS composite), red optical (NIRCam composite), and MIRI IR detections. A red square on the lower-resolution 15''$\times$15'' MIPS24 image indicates the size of the MIRI cutouts. The bottom row features larger 100''$\times$100'' cutouts in MIPS and PACS, with a red square highlighting the smaller MIPS24 field. {\it Center:} AGN-dominated SED fit (black line). Grey, red, and orange markers represent NIRCam+MIRI, Spitzer/MIPS, and Herschel photometry and upper limits. The blue line traces the unobscured galaxy host dominating the UV, while the purple line shows the obscured accretion disk dominating the optical. The dark and light red lines indicate hot and warm MBB dust emission from the AGN torus, and the magenta line shows the CLUMPY torus model \citep{nenkova08}. {\it Right:} Stellar-dominated SED fit (black line). The light and dark blue lines represent the low- and high-obscuration stellar populations dominating the UV and optical, respectively. The purple line indicates a minor AGN contribution. The light and dark red lines show ISM dust emission models from \citet{siebenmorgen07} and \citet{dl07} peaking at $\sim40~\mu$m. The grey lines, representing colder dust models from \citet{ce01}, \citet{dh02} and \citet{rieke09}, are inconsistent with the ALMA non-detection. {\it Bottom:} UDS-40579. Same panels as for JADES-191056. The cutouts show only the two PRIMER-MIRI bands, IRAC80, and MIPS24. This source is not detected at longer IR wavelengths. The best-fit SEDs are similar to those of JADES-191056, with the loose constraints from the IR upper limits allowing for both ISM- or AGN-dominated IR emission.}
\label{fig:LRD_lowz}
\end{figure*}

\subsection{SED modeling codes}
\label{s:fittingcodes}

In the following, we use two of the SED-fitting codes for LRDs described in detail in \citet{pg24a}: \textsc{prospector-AGN} and \textsc{Synthesizer-AGN}, both of which can model the observed SED with contributions from either an AGN or a galaxy continuum.

\textsc{Prospector-AGN}: uses a three-component model that includes the separate contributions from a galaxy, AGN accretion disk, and AGN torus. The stellar emission is modeled using Prospector \citep{leja19} with a parametric, delayed-$\tau$ SFH without dust attenuation. Consequently, it always has blue colors that dominate the rest-frame UV region of the SED.  The accretion disk emission consists of two components, one that follows the typical type 1 QSO template up to $\lambda=$5000~\AA\ and a power-law (f$_{\nu}=\lambda^{-\alpha}$) at longer wavelengths. We allow the spectral index of the power-law to vary between $\alpha=$0. and 0.59 (as in \citealt{temple21}) starting from a default value of $\alpha=$0.33 (as in \citealt{hernancaballero16}). The continuum is attenuated following a \citet{calzetti} law. Lastly, the torus dust emission is modeled in two possible ways, 1) using the subset of CLUMPY torus models from \citet{nenkova08} included in \textsc{Prospector}, and 2) using two modified black bodies (MBBs) that describe the combined emission from a hot and a warm dust component. The hot component has a fixed temperature of T$=1200$~K, while the temperature of the warm component is a fitting parameter that varies between T$=50$ and 500~K.

\textsc{Synthesizer-AGN}: uses a composite of two distinct stellar populations: a young and an evolved starburst with independent attenuations following a \citet{calzetti} law. The stellar emission is based on the \citet{bc03} models with a \citet{chabrier} IMF. Both SFHs follow delayed-$\tau$ models, with ages ranging from 1 Myr to the age of the Universe at the source's redshift. Nebular and dust emissions are also included, the former using Cloudy models (c23.01, \citealt{2023RNAAS...7..246G}) for a variety of gas temperatures and gas densities, the latter modeled using radiative transfer models of nuclear starbursts from \citet{siebenmorgen07}, as well as with templates from \citet{ce01,dh02,dl07,rieke09}. An AGN component is included using a type I QSO average spectrum extended to the far-IR and affected by its own (independent) \citet{calzetti} attenuation.

\subsection{Best-fit SEDs}

Figure~\ref{fig:LRD_lowz} shows the full SED and best-fit models for the two low-z LRDs computed with \textsc{prospector-AGN} and \textsc{synthesizer-AGN}. In addition to the HST and JWST photometry (grey points), the red, orange, and magenta markers indicate the mid and far-IR photometric detections and upper limits from Spitzer, Herschel, and ALMA. For JADES-191056, we fit the MIPS24 and PACS100 and 160 detections, while for UDS-40579, we use the most restrictive IR upper limits to illustrate the maximum dust emission allowed by the data.

{\bf JADES-191056:} The \textsc{Prospector-AGN} fit matches the observed SED well. This galaxy+AGN model suggests that the rest-frame UV up to $\lambda\sim4000$~\AA\ is dominated by a young ($\sim60$~Myr), unobscured, low-mass galaxy (\lmass$=8.2\pm0.1$; blue line), while the optical-to-NIR is dominated by an obscured accretion disk (A(V)$=3.8$~mag; purple line), with prominent \OIII\ and \Ha\ emission lines contributing to the broadband photometry in F277W and F356W. The MIRI photometry shows that the steep optical SED flattens at $\lambda>1~\mu$m, reaching a \nearmidir$=1.6$~mag color, similar to the median of F1800W-detected LRDs. The obscured accretion disk fits the near-to-mid IR SED well, with minimal contribution ($f\,^{\rm HD}_{3\mu m}<10$\%) from the hot-dust torus (T$=1250$~K; dark red). The mid-IR slope extends to $\lambda\sim5~\mu$m, where the strong MIPS24 detection suggests a transition into a spectral range dominated by warm-dust emission (T$=$167~K; light red). The best-fit IR SED, modeled with a single MBB, peaks at $\lambda\sim20~\mu$m, fitting the MIPS24 and PACS100 fluxes and being consistent with MIPS70, SPIRE, and ALMA upper limits. However, the model does not reproduce the PACS160 flux, possibly indicating that the AGN dust torus has a broader temperature distribution that a single-temperature MBB is unable to fit. A second fit using CLUMPY dust torus models \citep{nenkova08} (coral line) also fails to reproduce the IR SED but shows that the torus spectrum can be modeled as a combination of MBBs with temperatures ranging from hot dust (T$=$1250~K) to cooler dust peaking at $\lambda\sim40\,\mu$m. While we avoid more complex multi-component MBB fitting due to the limited and low-SNR data, we follow \citet{akins24} in presenting a dust model with a Gaussian temperature distribution (grey dashed line; mean 120 K, standard deviation 50 K) to demonstrate the broader range of temperatures implied by the data. We estimate a bolometric luminosity of $\log$(L$_{\rm bol}$[erg s$^{-1}$])=45.8 for the MBB model. This value is consistent with the obscured emission from the accretion disk in the UV-to-NIR range, $\log$(L$_{\rm bol,ABS}$[erg s$^{-1}$])=45.7, indicating an energy balance is possible with these obscured-AGN dominated models.

The \textsc{Synthesizer-AGN} model also provides a good fit, with the UV-to-NIR SED primarily driven by stellar emission. This is characterized by two young stellar populations ($\sim$2 and 45~Myr) with low and high obscurations (A(V)$=$0.8 and 4~mag; light and dark blue). An additional obscured AGN component (purple) contributes less than $\sim$15\% to the optical flux but dominates the $\lambda=2-3~\mu$m range. The total stellar mass (\lmass$=$9.8$\pm$0.1), dominated by the obscured stellar population, is $\sim2$~dex higher than the results with \textsc{Prospector-AGN}, where stars dominate the UV emission but contribute negligibly to the optical emission; however, it remains within the range of typical massive galaxies at this redshift. The mid-IR emission is well reproduced by \citet{siebenmorgen07} and \citet{dl07} models, peaking at $\lambda\sim40\,\mu$m (T$\sim70$~K), similar to star-forming and photodissociation regions (e.g., \citealt{elbaz11}) and consistent with the ALMA 1.1 mm non-detection (and low SNR SPIRE data points, \citealt{magnelli13}). These dust models have a bolometric luminosity of $\log$(L$_{\rm bol}$[erg s$^{-1}$])$=46.0$ that is fully consistent (within 0.1~dex) with obscured stellar emission in the UV-to-NIR range. In contrast, alternative fits using standard dust models for lower-redshift dusty galaxies (\citealt{ce01}; \citealt{dh02}; \citealt{rieke09}; solid, dashed and dash-doted grey lines) predict cooler dust temperatures peaking around $\lambda\sim90~\mu$m, which are inconsistent with the ALMA non-detection and predict even larger bolometric luminosities. In particular, the steep flux increase from MIRI to MIPS24 at $\lambda\gtrsim6~\mu$m suggests the presence of strong PAH emission lines, highlighting the potential for spectroscopic MIRI follow-up to detect the prominent 3.3~$\mu$m PAH line. 

JADES-191056 highlights the critical role of low-redshift LRDs in constraining dust emission peaks and improving estimates of dust masses and temperatures. However, substantial degeneracies persist due to uncertainty as to whether IR emission originates from the galaxy’s extended ISM or the compact AGN region. If the emission arises from the ISM, hot or warm dust peaking at $10-40~\mu$m may dominate the IR luminosity, while colder dust peaking beyond $>50$~$\mu$m could contribute significantly to the dust mass with a lesser impact on the luminosity. The \citet{dl07} fit yields a dust mass estimate of M$_\mathrm{dust}=10^{7.1\pm0.3}$~M$_{\odot}$, assuming a dust-to-gas ratio of $10^{-2.8}$, as derived for a $z\sim3$ galaxy with \lmass$\sim10$ from \citet{2017MNRAS.471.3152P}. This corresponds to a dust-to-star mass ratio of M$_\mathrm{dust}$/M$_{\star}=0.0022$, consistent with reported values for galaxies of similar masses and redshifts \citep[e.g.,][]{2014A&A...562A..30S,2017MNRAS.465...54C,2020A&A...644A.144D}. Alternatively, if the IR emission originates from a hot-dust-deficient AGN with a broader range of warm dust temperatures, luminosities of $\log$(L$_{\rm bol}$[erg s$^{-1}$])$=46.0$ could correspond to lower dust masses. The value depends on whether the AGN hosts a classical dust torus ($\lesssim10$~pc) or a more extended distribution (a few 100~pc), indicative of an early formation stage. For example, \citet{li24} proposed that LRDs might feature a more spherical distribution of gas and dust with densities of $\rho\sim10–10^{3}$~cm$^{-3}$, capable of producing colder dust temperatures (as in the polar regions of the torus; \citealt{ramosalmeida17}) and high luminosities ($\log$(L$_{\rm bol}$[erg s$^{-1}$])$\sim45.7$) with modest dust masses (M$_\mathrm{dust}=10^{4-5}$~M$_{\odot}$) while remaining consistent with the flattening of the IR SED observed at $1–5~\mu$m.

The IR-based luminosity and ISM dust mass derived for JADES-191056 are significantly higher than the median values reported in \citet{casey24} for a large sample of LRDs, estimated using empirical relations for UV luminosities and dust attenuations. However, our best-fit SED agrees with the 95th percentile confidence region of their IR SED and is consistent with the maximal IR SED stack in \citet{akins24}, which has a higher luminosity ($\log$(L$_{\rm bol}$[erg s$^{-1}$])=45.9). Scaling the SED of JADES-191056 to $z=6$ suggests a PACS100 flux of $\sim$450~$\mu$Jy, comparable to the 5$\sigma$ detection limit of the \citet{akins24} stack (400~$\mu$Jy). Note also that JADES-191056 is among the $\sim$ 5\% most luminous LRDs in the sample (M$_{1\mu m}=-22.8$~mag, $\sim$1.3~mag brighter than the median), indicating that most LRDs likely exhibit lower near-IR and total IR luminosities, falling below the PACS100/160 upper limits if their IR SEDs are similar.

{\bf UDS-40579}: This LRD exhibits the reddest UV-to-NIR colors in the sample, with most NIRCam bands probing a steep, red SED. Only a few HST/ACS and WFC3 bands capture a faint blue UV emission at $\lambda_{\rm rest}\lesssim2500$~\AA. The \textsc{prospector-AGN} fit
indicates an obscured accretion disk (A(V)$=4$~mag) dominates the SED, with a minor contribution from a low-mass blue galaxy (\lmass$=8.1\pm0.1$) in the rest-UV. Unlike JADES-191056, the transition from galaxy to accretion-disk dominated continuum occurs at shorter wavelengths ($\lambda\sim3000$~\AA), leading to redder UV-to-NIR colors. However, the near-to-mid IR color is relatively small (\nearmidir$=1.1$~mag; F770W$-$F1800W$\sim0$~mag), and the strong mid-IR flattening requires no additional hot dust to fit the SED. The Spitzer/Herschel upper limits allow for a potential warm-dust component dominating at $\lambda_{\rm rest}\gtrsim6~\mu$m. However, the lack of stringent constraints makes it difficult to determine the exact IR peak, which might shift to longer wavelengths if, for instance, the MIPS24 upper limit were excluded. These results agree well with \citet{wang24_lrd}. They find a slightly higher stellar mass (\lmass$=9.25^{+0.15}_{-0.11}$) likely due to their galaxy component dominating the flux up to $\lambda_{\rm rest}\sim4500$~\AA.

The \textsc{synthesizer-AGN} model suggests a similar configuration to JADES-191056, with low- and high-obscuration stellar populations (A(V)$=1.5$ and 3.8~mag) dominating the SED up to $\lambda\sim1~\mu$m, and the obscured AGN dominating at $\lambda\sim2-3~\mu$m to account for SED flattening. The redder optical colors are reproduced by a brighter, obscured stellar component, leading to a higher stellar mass (\lmass$=10.3\pm0.2$) than the other LRD. The IR models suggest that an ISM-dust-dominated solution is consistent with the broader constraints set by the Spitzer/Herschel upper limits. Even assuming no ALMA detection within limits similar to the JADES source, the \citet{siebenmorgen07} model, with higher dust temperatures peaking at $\lambda\sim50~\mu$m, remains compatible with the non-detection. In this case, the fits to the UV-to-NIR fluxes imply an absorbed energy that makes the galaxy extremely luminous, with $\log$(L$_{\rm bol}$[erg s$^{-1}$])$=46.7$. The dust emission fits provide a similarly high luminosity of $\log$(L$_{\rm bol}$[erg s$^{-1}$])$=46.3$ and a large dust mass of M$_\mathrm{dust}=10^{9.0\pm0.7}$~M$_\odot$, although these should be regarded as an upper limit, as we do not detect the source in the far-IR and we do not have data to probe the sub-mm range for this source.

\subsection{Discussion}

The analysis above shows the value of bright low-redshift LRDs in finding the peak of IR emission, enabling more accurate estimates of total dust content and characteristic temperatures. However, it also highlights that the SED degeneracies between galaxy and AGN in the UV and optical extend into the near- and mid-IR. For instance, the pronounced upturn at $\lambda\sim6\,\mu$m in JADES-191056 could result from PAH emission from the galaxy's ISM or dust at T~$\sim170$~K associated with the AGN. Similarly, the IR emission peak at $\sim30-40\mu$m, likely arising from dust spanning a wide temperature range (T~$\sim100-70$~K), might originate from the galaxy-wide ISM or an extended nuclear AGN region. These degeneracies lead to significantly different estimates of stellar and dust masses depending on whether galaxy- or AGN-dominated models are assumed, implying different natures for LRDs.

Are LRDs compact, massive dusty starbursts or small, low-mass, low-extinction galaxies with a central obscured AGN? If the latter, where is the dust coming from, and why is the host galaxy blue? Galaxy formation at high redshift is more gas-rich and chaotic, which can trigger intense episodes of ``wet compaction" (\citealt{dekel13b}; \citealt{wellons15}) that cause strong starburst, leading to higher dust and stellar concentrations (\citealt{zolotov15}; \citealt{tacchella16}). These processes would favor the growth of a supermassive black hole (SMBH) within a central dusty region, with continuous inflows and outflows. However, this scenario should coexist with star formation, as seen in compact, dusty SFGs at $z\sim2-3$, which show obscured nuclear bursts (\citealt{barro17}; \citealt{tadaki17b}) and a high incidence of AGNs \citep{kocevski17}. It would be surprising if the dust were confined only to the central AGN region without being associated with galaxy-wide star formation, which would exhibit lower dust temperatures and higher dust masses. Nevertheless, if LRDs are predominantly massive dusty galaxies, their stellar masses and densities are unexpectedly large for their volume densities (\citealt{akins24}; \citealt{leung24}) and their large attenuations require unusually gray attenuation laws (\citealt{barro24,pg24a}), which have to remain surprisingly consistent over a wide redshift range ($4 \lesssim z < 9$) to produce the narrow distribution of flat UV SEDs seen in LRDs.

\subsection{Prospects for explaining the origin of the IR emission}

Additional data are needed to resolve the existing degeneracies and determine where LRDs fit within the galaxy or AGN edge scenarios. Deeper JWST spectroscopy of the rest-frame optical continuum, searching for stellar absorption lines or continuum breaks, could show if LRDs have a dominant stellar component. However, breaking IR SED degeneracies to reveal the origin of dust emission presents further challenges, as the peak of the IR emission lies between the observable MIRI range and the capabilities of current ALMA observations (e.g., \citealt{casey24}). This is why LRDs at lower redshifts provide a unique observational window to probe deeper into the rest-frame near- to mid-IR wavelengths, offering weak detections or tighter constraints from complementary mid- to far-IR or sub-mm data. For example, MIRI spectroscopy of low-z LRDs could identify the PAH emission line at 3.3~$\mu$m, typically associated with ISM dust, which is weak or absent near AGNs (e.g., \citealt{buiten24}). This could help differentiate between galaxy- and AGN-driven contributions to the IR emission. Similarly, deep MIRI spectroscopy could detect stellar absorption lines in the NIR continuum, such as the CO overtones \citep{bianchin24}, providing novel constraints that complement UV and optical spectroscopy, where such lines remain undetected.

Another significant challenge in understanding the IR emission of LRDs stems from the limited depth of current MIRI data at longer wavelengths ($\gtrsim15\mu$m), resulting in low detection fractions (7\%) and weak upper limits on IR colors (\nearmidir$<2.3$~mag). These limits still allow for dust emission fractions as high as 70–80\%. Notably, increasing the depth of the PRIMER F1800W imaging by $\sim1$~mag would reduce the median upper limit of non-detections to a color (\nearmidir$\lesssim1$~mag) that could be modeled without invoking any dust emission. This improvement would provide much tighter constraints on the dust content of nearly half the F770W-detected LRDs, which constitute 77\% of the sample to F444W$=27$~mag.

\section{Summary}
\label{s:summary}

We present a new photometric selection for Little Red Dots (LRDs) designed to capture their broad range in redshifts and rest-frame UV-to-NIR colors, offering a more comprehensive view of their properties. Using this method, we identify 248 LRDs with F444W~$<27$~mag across a combined area of 263~arcmin$^{2}$ in the JADES, PRIMER-COSMOS, and PRIMER-UDS fields with MIRI coverage. MIRI photometry is crucial for extending the NIRCam SED coverage from rest-frame UV and optical into the near- and mid-IR. These spectral regions are essential for identifying bluer LRDs, minimizing the risk of emission line contamination, and probing IR dust emission. Our key findings are as follows:

\begin{itemize}
\item We compare the new selection with previous samples of LRDs from \citet{kokorev24} and \citet{kocevski24}, identified with typical photometric criteria. These samples select a similar number of LRDs but only have $\sim50$\% in common. The new selection recovers 93\% of the LRDs in either of these samples and finds an additional $\times1.7$ more candidates with a total number density of 0.94 arcmin$^{-2}$.

\item LRDs have a median magnitude, color, and redshift of F444W~$=25.9^{+0.7}_{-1.1}$~mag, F277W$-$F444W~$=1.0^{+0.7}_{-0.6}$~mag, and $z=5.5^{+1.9}_{-0.8}$. The lower median redshift relative to previous LRDs samples ($z\sim6.5$) is driven by new identifications at $z<5.5$ with bluer colors (F277W$-$F444W~$<1$~mag) likely caused by strong \OIII\ emission in F277W.

\item The majority of LRDs are detected in MIRI/F770W, 94\% (77\%) at F444W~$<26$ (27)~mag, while only 7\% (17) are detected in F1800W. F1800W detections are typically among the most luminous LRDs with M$_{1 \mu m}<-22$~mag. Their median color  F770W$-$F1800W$=1.4^{+0.4}_{-0.6}$~mag is $\sim$1.5~mag bluer than a typical, dust-obscured QSO at $z\gtrsim5$, in agreement with previous works.

\item We use a rest-frame near-to-mid IR color as a tracer of dust emission. F1800W-detected LRDs have a median color of \nearmidir$=1.5^{+1.1}_{-0.7}$~mag, with significant scatter, indicating diverse dust emission properties across the population. 

\item We compare the \nearmidir color to predictions from galaxy- and AGN-dominated LRD models. Only $\sim20$\% show colors \nearmidir$\lesssim1$~mag consistent with negligible dust emission, while the majority ($\sim50$\%) exhibit dust fractions as high as f\,$^{\rm ISM}_{3\mu m}=0.7$ or f\,$^{\rm HD}_{3\mu m}=0.8$. These are still lower than in dust-obscured QSOs (f\,$^{\rm HD}_{\rm QSO}\gtrsim0.2$). The limited coverage of the IR SED prevents a definitive determination of whether the dust emission originates from the galaxy's ISM or an AGN torus.

\item LRDs exhibit a median rest-frame UV-to-NIR color of \uvnir$=2.5^{+0.8}_{-0.6}$~mag. The $\sim2$~mag scatter reflects the diverse intrinsic properties of LRDs. F1800W-detected LRDs exhibit a trend of redder \nearmidir\ (indicating stronger IR emission) with bluer \uvnir, implying that the SEDs of the bluest LRDs may resemble those of typical unobscured QSOs.

\item We identify an IR-bright LRD at $z_{\rm spec}=3.1386$ with UV-to-NIR colors typical of the LRD population. We fit its IR SED, including data from 7 MIRI bands (up to F2100W), Spitzer/MIPS24, Herschel PACS100 and PACS160, and an ALMA 1.1mm non-detection. The IR SED fit shows a steep reddening at $\lambda_{\rm rest}>6~\mu$m and a peak at $\sim30-40~\mu$m which may indicate the first direct detection of warm dust emission in a LRD.

\end{itemize}

\section*{Acknowledgments}
We thank G. Brammer and the Dawn JWST Archive (DJA) for making high-level data products publicly available. DJA is an initiative of the Cosmic Dawn Center, which is funded by the Danish National Research Foundation un- der grant No. 140. PGP-G acknowledges support from grant PID2022-139567NB-I00 funded by Spanish Ministerio de Ciencia, Innovaci\'on y Universidades MCIU/AEI/10.13039/501100011033,
FEDER {\it Una manera de hacer
Europa}. JSD acknowledges the support of the Royal Society via the award of a Royal Society Research Professorship. This work has made use of the Rainbow Cosmological Surveys Database, which is operated by the Centro de Astrobiología (CAB), CSIC-INTA, partnered with the University of California Observatories at Santa Cruz (UCO/Lick, UCSC). 

\software{Astropy \citep{astropy}, EAZY \citep{eazy}, GALFIT  \citep{galfit}, matplotlib \citep{matplotlib}, NumPy \citep{numpy}, PZETA (\citealt{pg08}), Prospector (\citealt{leja19} \citealt{johnson21}), Rainbow pipeline (\citealt{pg05,pg08}, \citealt{barro11b}), SExtractor \citep{sex}, Synthesizer (\citealt{pg05,pg08b})}

\bibliography{referencias}

\begin{appendix}
\section{MIPS24 photometry of JADES-191056}
\label{s:mips_flux}

\begin{figure}%[htp!]%[ht!]
\centering
\includegraphics[width=18cm,angle=0]{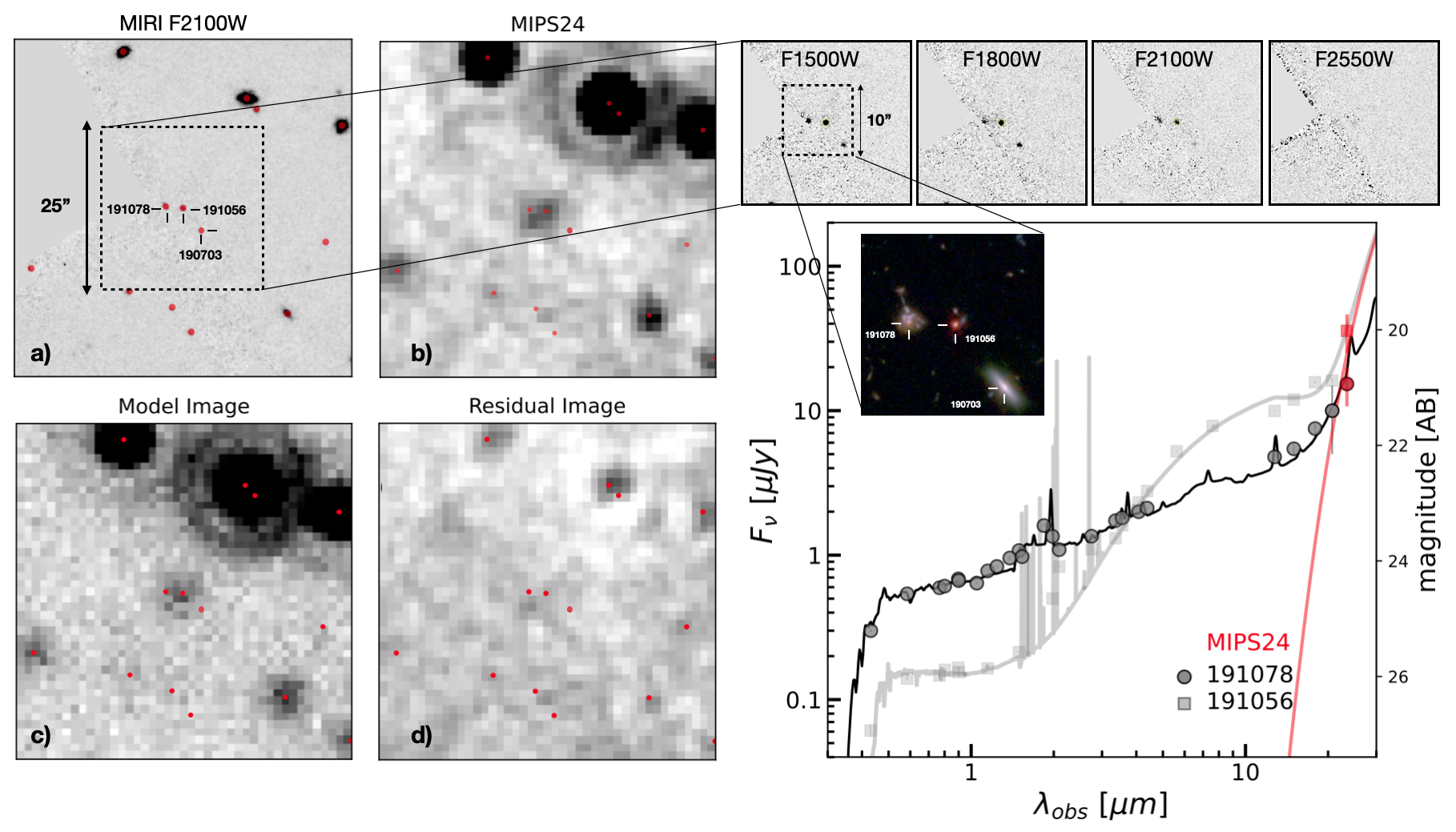}
\caption{MIPS24 photometry of JADES-191056. {\it Left}: Panels a) and b) show 51''$\times$51'' cutouts of the MIRI/F2100W and Spitzer/MIPS24 images around JADES-191056. The high-resolution MIRI image shows the primary source and two nearby companions. The dashed line indicates the 25''$\times$25'' zoom-in region shown in other MIRI images on the right. The MIP24 image shows the position of multiple sources detected in F2100W (red dots) and highlights the blending of JADES-191056 and JADES-191078. Panels on the bottom describe the PSF-matching photometry: c) shows the best-fit model image created from PSFs located at the position of the F2100W sources; d) shows the clean residual after subtracting the model from the primary image.  {\it Right:} The top 4 panels show the 25''$\times$25'' zoom-in region in the longest wavelength MIRI bands (note the non-detection in F2550W). The dashed line shows a 10''$\times$10'' zoom in matching the NIRCam composite (F150W+F277W+F444W) image below that highlights the different appearances of JADES-191056 and its companions. The bottom panel shows the best-fit SEDs of JADES-191056 (light grey) and its closest companion JADES-191078. Their respective MIPS24 fluxes derived from the PSF-photometry are shown in red.}
\label{fig:mips24_deblending}
\end{figure}

Here we describe in more detail the MIPS24 photometric measurement for JADES-191056 to show that it is robust against contamination by nearby sources. The high resolution images show that the primary LRD JADES-191056 has a nearby galaxy companion, JADES-191078, at d$=$2.61'' which is also IR bright, detected in MIRI photometry up to F2100W. The companion galaxy has a photometric redshift of z$=$2.91 which indicates that MIRI is probing the rest-frame near-to-far IR of the wavelengths and, most likely, dust emission. Since the two sources are closer than the nominal spatial resolution of the MIPS24 data (FWHM$=$5'') they appear blended in the low-resolution data. There is a third possible companion (JADES-190703) at a larger distance, d$=$4.74'', with very low z$=$0.51, that is also detected in MIRI up to F1800W. However, it becomes very faint in F2100W and it is far enough that it is unlikely that it can contribute to the blend in MIPS24.

To estimate deconvolved MIPS24 photometry for the LRD and its companion we use a PSF fitting method that relies on the higher spatial resolution data from MIRI to determine the central position of the PSF in the MIPS24 image. This photometric procedure is very similar to the default approach to computing MIPS photometry (see e.g., \citealt{pg05}; \citealt{pg08}) but it adds position priors to indicate that the source is blended.  A similar approach has also been applied to compute IRAC photometry using HST or JWST priors (e.g. TPHOT; \citealt{tphot}). Since the difference in spatial resolution between MIRI F2100W (FWHM$=$0.67'') and MIPS24 is larger than $\times$7, a PSF fitting method provides similar results to a template fitting approach based on the F2100W imaging with convolution kernel to create a MIPS24 model. The fitting is computed using Astropy/photutils PSF fitting routines. We use the MIPS24 PSF computed from a stack of bright isolated sources as described in \citet{pg05}. The integrated, blended, photometry in the MIPS24 image yields a flux of f$_{\nu}$$=$53.7$\pm$5.5$\mu$Jy. The PSF fitting photometry returns individual fluxes for the LRD, f$_{\nu}$$=$35.7$\pm$3.8$\mu$Jy, and the companion, f$_{\nu}$$=$12.3$\pm$1.7$\mu$Jy. This implies roughly a 70\% to 30\% breakout which seems consistent with the spatial distribution of the sources showing that JADES-191056 is closer to the centroid of the MIPS24 detection.

The 4 panels of the left side of Figure~\ref{fig:mips24_deblending} show 50''$\times$50'' cutouts around JADES-191056 in a) MIRI F2100W, b) MIPS24, c) the MIPS24 model computed by fitting the original MIPS24 image with photutils, d) the residual image resulting from subtracting the original and model images. The red circles indicate the prior locations based on the brightest ($>$10$\sigma$ detections in F2100W). The top 4 panels on the right side show smaller 25''$\times$25'' cutouts around JADES-191056 in MIRI F1500W, F1800W, F2100W, and F2550W. The primary LRD and, especially its companion are both relatively close to the edge of the mosaic. As a result, due to the survey design, the latter has no data in F560W, F770W, and F1000W and JADES-191056 has no data in F1000W. Neither are detected in the longest wavelength data F2550W due to the shallow limiting magnitude of the data in F2550W($5\sigma$)$=$20.8) and the proximity to the edge of the mosaic. The bottom right panel shows the observed SEDs and best-fit models for the LRD (squares) and its companion (circles). The AGN model for the LRD is described in \S~\ref{s:lowz_LRDs}. For the galaxy, we use the HST and JWST photometry from the JADES DR2 catalog and we measure the MIRI fluxes following the same methods described in \S~\ref{s:data}. For the SED fitting we use Prospector with the default modeling assumptions and including the effects of dust emission.

The MIRI photometry shows that the LRD is brighter than the galaxy companion in all bands by $\sim$0.7~mag, although the gap narrows to $\sim$0.3~mag in F2100W. The overall trends in both SEDs is consistent with the larger share of the MIPS24 flux for the LRD found above. The SED models show that, due to the relatively close redshifts of the LRD and its companion, their IR-SEDs appear to rise steeply in the MIRI to MIPS24 range as they enter the wavelength range dominated by PAHs $\lambda\gtrsim6\mu$m. While the SED of the LRD is always above the companion, a possible concern would be if the IR SED of the galaxy rises more steeply than the current fit shows, and thus a larger share of the MIPS24 flux comes from the galaxy. However, this scenario appears inconsistent with the spatial distribution of the sources and it would imply much larger IR luminosities for the galaxy which is still not detected in the relatively deep Herschel data or the ALMA map.

Following this analysis, we conclude that JADES-191056 is likely the primary emission source in the MIPS24 detection. The fluxes for this source used throughout the paper are based on the measurements described above.

\section{Table, Images and SEDs of the F1800W detections.}
\label{s:f1800w_seds}
\begin{center}
\begin{table}[h!]
%\centering
\small
\begin{tabular}{cccrrrrrr}
\toprule
ID    &    RA        &     DEC    & z$_{\rm phot}$ & z$_{\rm spec}$   & F770W & F1800W  & \uvnir &  \nearmidir \\
      &    [deg]     &    [deg]   &       &        & [mag] & [mag]   &  [mag] & [mag]\\
\hline
50988$^{e}$ &  150.17593817 & 2.43242976 & 4.58 & - & 22.80 & 21.40 & 3.51 & 1.41 \\ 
50808$^{p}$ &  150.20216272 & 2.43197607 & 4.69 & - & 25.55 & 22.52 & 1.99 & 2.61 \\
40053$^{m}$ &  150.16199469 & 2.37607598 & 5.17 & - & 24.26 & 21.39 & 2.61 & 3.74 \\
31661$^{n}$ &  150.05965530 & 2.29343489 & 5.65 & - & 23.95 & 22.45 & 2.24 & 1.98 \\
22702$^{c}$ &  150.06375097 & 2.24679796 & 5.44 & - & 22.82 & 23.20 & 3.80 & 0.13 \\ 
13074$^{d}$ &  150.07166522 & 2.20039773 & 5.20 & - & 22.94 & 21.90 & 3.49 & 1.26 \\ 
11367$^{k}$ &  150.09827909 & 2.19218392 & 3.81 & - & 24.60 & 23.18 & 2.72 & 1.59 \\ 
\hline                                                                             
61263$^{o}$ &  34.47092455 & -5.10678699 & 3.63 & - & 24.36 & 22.89 & 2.21 & 2.62 \\  
47101$^{b}$ &  34.46034794 & -5.13157941 & 4.41 & - & 21.32 & 19.35 & 4.11 & 2.78 \\  
32237$^{f}$ &  34.41984016 & -5.15422798 & 4.63 & - & 23.67 & 22.65 & 3.22 & 0.70 \\
40579$^{a\dagger\dagger}$ &  34.24419037 & -5.24583353 & 3.56 & 3.1034 & 21.25 & 22.20 & 4.82 & 1.12 \\
38220$^{g}$ &  34.23088949 & -5.24952987 & 4.34 & - & 23.75 & 22.10 & 3.25 & 1.97 \\
11501$^{l}$ &  34.42091084 & -5.29617799 & 4.68 & - & 23.44 & 22.66 & 2.63 & 0.95 \\
\hline
57356$^{j\dagger}$ &  53.11531605 & -27.85921725 & 4.27 & - & 24.39 & 22.86 & 2.90 & 1.87 \\
191056$^{h}$ &  53.16481198 & -27.83696943 & 3.10 & 3.1386 & 21.66 & 20.91 & 3.09 & 1.64 \\
193912$^{q}$ &  53.14259358 & -27.82659282 & 3.69 & - & 23.13 & 22.37 & 1.92 & 1.26 \\
204851$^{i\dagger}$ &  53.13859362 & -27.79025334 & 5.42 & - & 23.70 & 23.42 & 3.01 & 0.77 \\
\hline
\end{tabular}
\caption{Table with coordinates, observed MIRI magnitudes, and rest-frame UV-to-NIR and near-to-mid IR colors for the 17 F1800W detected LRDs in the PRIMER-COSMOS, PRIMER-UDS and JADES fields. The letters indicate the LRDs in the central panel of Figure~\ref{fig:ircolors}. The LRDs in common with \citet{pg24a} and \citet{wang24_lrd} are indicated with $\dagger$ and $\dagger\dagger$.}
\label{tab:lrd18}
\end{table}
\end{center}

\begin{figure}%[htp!]%[ht!]
\centering
\includegraphics[width=18cm,angle=0]{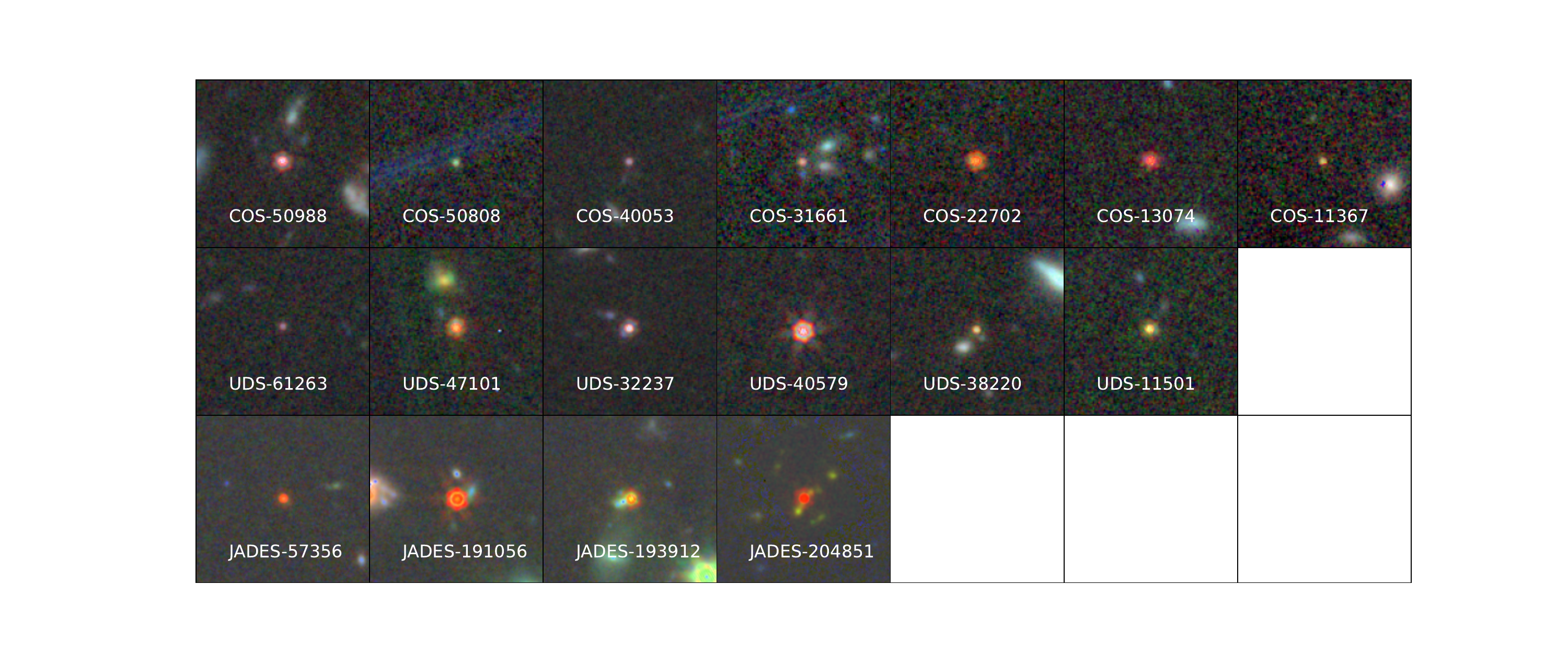}\\[-10mm]
\includegraphics[width=18cm,angle=0]{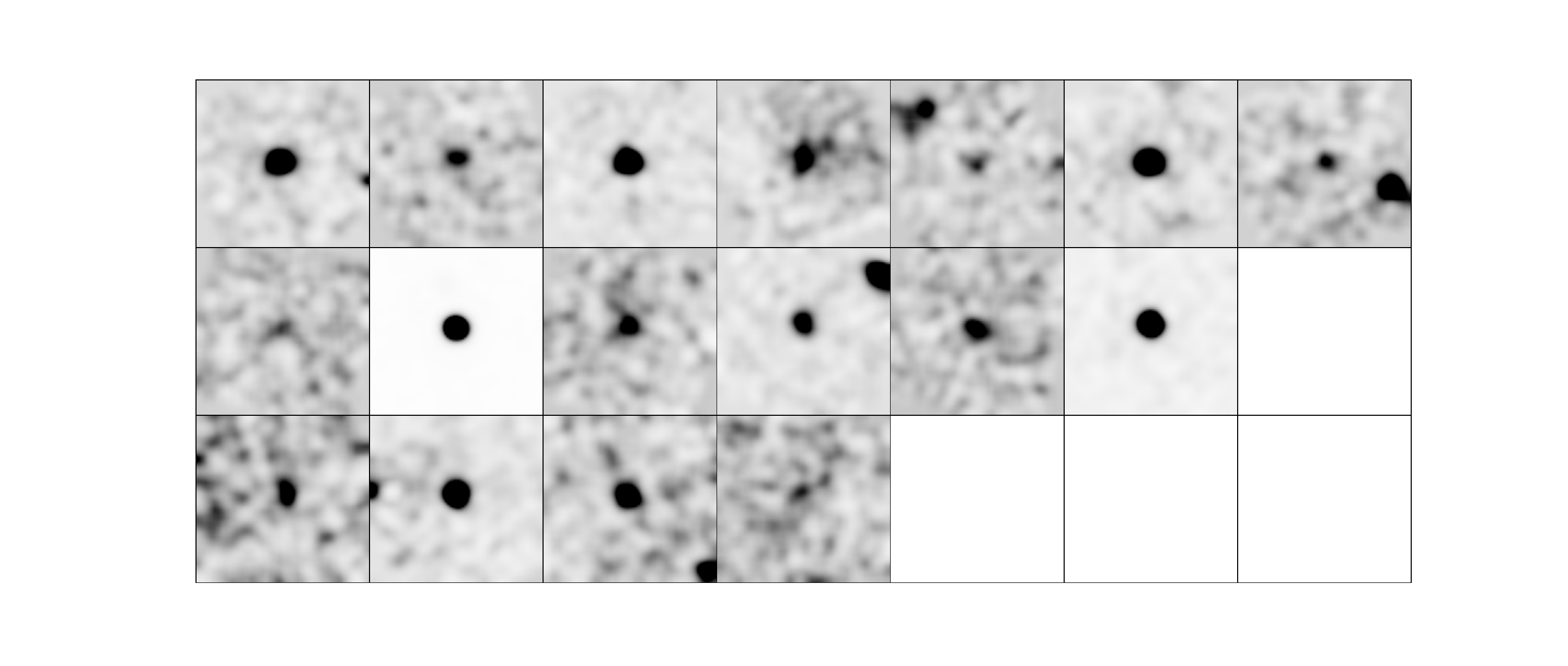}\\[-10mm]
\includegraphics[width=18cm,angle=0]{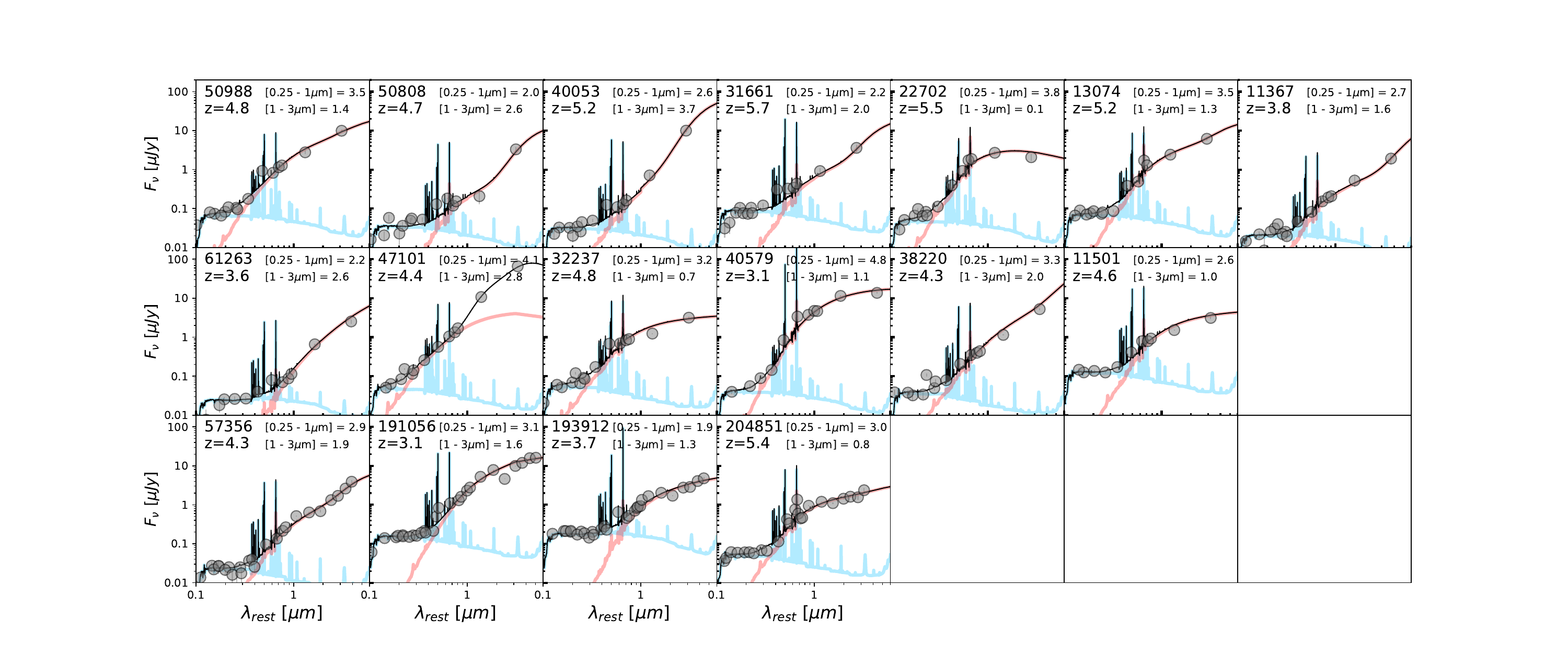}
\caption{{\it Top:}  5''$\times$5'' NIRCam color (F150W+F277W+F444W) cutouts of the F1800W-detected LRDs. Each row shows 7, 5 and 4 sources in PRIMER-COSMOS, UDS and JADES, respectively. {\it Middle:} 5''$\times$5'' MIRI/F1800W cutouts of the same LRDs smoothed with a 3-pixel Gaussian kernel to increase the contrast. {\it Bottom:} Best-fit SEDs of the same LRDs with the two-component AGN-based model of Prospector-AGN.}
\label{fig:F1800W_detections_properties}
\end{figure}

\end{appendix}

\end{document}